\pdfoutput=1

\documentclass[11pt,twoside,a4paper,cmspaper,final,collab]{cms-tdr}
\def\svnVersion{465761P}\def\cmsCernNoTag{CERN-EP-2017-336}\def\cmsCernDate{\today}\def\cmsMessage{Published in Physics Letters B as \href{http://dx.doi.org/10.1016/j.physletb.2018.05.062}{\doi{10.1016/j.physletb.2018.05.062.}}}

\begin{document}\cmsNoteHeader{SUS-16-048}

\hyphenation{had-ron-i-za-tion}
\hyphenation{cal-or-i-me-ter}
\hyphenation{de-vices}
\RCS$Revision: 463155 $
\RCS$HeadURL: svn+ssh://svn.cern.ch/reps/tdr2/papers/SUS-16-048/trunk/SUS-16-048.tex $
\RCS$Id: SUS-16-048.tex 463155 2018-06-04 11:29:30Z bschneid $
\newlength\cmsFigWidth
\ifthenelse{\boolean{cms@external}}{\setlength\cmsFigWidth{0.98\columnwidth}}{\setlength\cmsFigWidth{0.70\textwidth}}
\newlength\cmsFigWidthTwo
\ifthenelse{\boolean{cms@external}}{\setlength\cmsFigWidthTwo{0.98\columnwidth}}{\setlength\cmsFigWidthTwo{0.32\textwidth}}
\ifthenelse{\boolean{cms@external}}{\providecommand{\cmsLeft}{upper\xspace}}{\providecommand{\cmsLeft}{left\xspace}}
\ifthenelse{\boolean{cms@external}}{\providecommand{\cmsRight}{lower\xspace}}{\providecommand{\cmsRight}{right\xspace}}
\newlength\cmsTabSkip\setlength{\cmsTabSkip}{1.8ex}
\ifthenelse{\boolean{cms@external}}{\newcommand{\cmsTabResize}[1]{#1}}{\newcommand{\cmsTabResize}[1]{\resizebox{\textwidth}{!}{#1}}}

\providecommand{\NA}{\text{---}\xspace}
\newcommand{\Wjets}{{\ensuremath{\PW\text{+jets}}}\xspace}
\newcommand{\Zjets}{{\ensuremath{\cPZ\text{+jets}}}\xspace}
\newcommand{\MT}{\ensuremath{M_{\mathrm{T}}}\xspace}
\newcommand{\Mll}{\ensuremath{M(\ell\ell)}\xspace}

\cmsNoteHeader{SUS-16-048}
  \title{Search for new physics in events with two soft oppositely charged leptons and missing transverse momentum in proton-proton collisions at $\sqrt{s}=13\TeV$}

\date{\today}

\abstract{A search is presented for new physics in events with two low-momentum,
oppositely charged leptons (electrons or muons) and missing transverse
  momentum in proton-proton collisions at a centre-of-mass energy of 13\TeV. The
  data collected using the CMS detector at the LHC correspond to an integrated
  luminosity of 35.9\fbinv.  The observed event yields are
  consistent with the expectations from the standard model. The results are
  interpreted in terms of pair production of charginos and neutralinos
  ($\PSGcpmDo$ and $\PSGczDt$) with nearly
  degenerate masses, as expected in natural supersymmetry models with light
  higgsinos, as well
  as in terms of the pair production of top squarks (\PSQt), when
  the lightest neutralino and the top squark have similar masses. At 95\%
  confidence level, wino-like $\PSGcpmDo$/$\PSGczDt$ masses
  are excluded up to 230\GeV for a mass difference
  of 20\GeV relative to the lightest neutralino.
  In the higgsino-like model, masses are excluded up to 168\GeV for the same
  mass difference.
 For \PSQt pair
production, top squark masses up to 450\GeV are
excluded for a mass difference of 40\GeV relative to the lightest
neutralino.}

\hypersetup{%
pdfauthor={CMS Collaboration},%
pdftitle={Search for new physics in events with two soft oppositely charged leptons and missing transverse momentum in proton-proton collisions at sqrt(s)=13 TeV},%
pdfsubject={CMS},%
pdfkeywords={CMS, physics, SUSY, compressed, leptons, missing energy, 13 TeV}}

\maketitle

\section{Introduction}\label{sec:introduction}

Supersymmetry (SUSY)~\cite{SUSY0,SUSY1,SUSY2,SUSY3,SUSY4} is a widely considered extension
of the standard model (SM) of particle physics, as it can provide solutions to
several open questions in the SM, in particular those related to the hierarchy
problem~\cite{LightStop1,Witten:1981nf,Dimopoulos:1981zb} and the nature of dark
matter.  SUSY predicts superpartners of SM particles whose spins differ by
one-half unit with respect to their SM partners.  In $R$-parity conserving
models~\cite{Farrar:1978xj}, SUSY particles are pair-produced and their decay
chains end in the stable, lightest SUSY particle (LSP), which in many models
corresponds to the lightest neutralino (\PSGczDo). A stable LSP would escape
undetected, yielding a characteristic signature of a large magnitude of
missing transverse momentum (\ptmiss) in collisions at the CERN LHC. As a stable, neutral and
weakly interacting particle, the neutralino matches the properties required of a
dark matter candidate~\cite{Patrignani:2016xqp}.

The absence of SUSY signals in previous experiments, as well as at the LHC, can
be interpreted as an indication that SUSY particles have very large mass,
leading to the expectation that SUSY events have large visible energy and
momentum. As a result, the many searches that yield the most stringent limits on
the masses of the SUSY particles are based on events with large \ptmiss and
energetic final-state objects such as leptons and jets.  Another interpretation
for the absence of a SUSY signal is that the SUSY particles are in a part of
the parameter space that is not easily accessible.  One such scenario, where
previously mentioned searches would not be sensitive, is where the mass spectrum
is compressed, i.e. the mass splitting between the produced SUSY particles and
the LSP is small.  When the mass splittings between SUSY particles are small,
the visible energy in the event, and also potentially the \ptmiss, is relatively
low, which motivates searches in events with low-momentum objects.

Compressed mass spectra arise in several SUSY models, including natural SUSY,
\ie SUSY models that solve the hierarchy problem with little fine tuning. It has
been pointed out in several studies, for example in
Refs.~\cite{LightStop1,LightStop2,Dimopoulos:1981zb,Witten:1981nf,Dine:1981za,Dimopoulos:1981au,Sakai:1981gr,Kaul:1981hi},
that naturalness imposes constraints on the masses of higgsinos, top squarks,
and gluinos. Natural SUSY is generally considered to require at least one
coloured SUSY particle of mass below approximately one \TeV.  Further, it is
often assumed that this particle is the top squark (\PSQt). More recently,
however, the hypothesis of natural SUSY requiring a light top squark has been
disputed as arising from oversimplified
assumptions~\cite{Casas:2014eca,Baer:2014ica,Mustafayev:2014lqa}. Irrespective
of the top squark, higgsinos remain a complementary window to natural SUSY as
they are generally expected to be light. As pointed out in
Refs.~\cite{Giudice:2010wb,Baer:2014kya,Han:2013usa,Han:2014kaa}, light
higgsinos are likely to have a compressed mass spectrum, potentially leading to
signatures with soft leptons and moderate \ptmiss.
Thus far, the most sensitive
searches in this model have been carried out by experiments at
LEP~\cite{Heister:2002mn,Abdallah:2003xe} and ATLAS~\cite{Aaboud:2017leg}.
The LEP experiments excluded $\PSGcpmDo$
masses up to 103.5\GeV for a mass splitting between the $\PSGcpmDo$ and $\PSGczDo$ of
at least 3\GeV.

The search described in this letter is designed for neutralinos and charginos,
which are collectively referred to as ``electroweakinos'', in a model where
these electroweakinos form a compressed mass
spectrum~\cite{Giudice:2010wb,Han:2013usa,Schwaller:2013baa,Han:2014kaa}.  Two
models are considered where the electroweakinos are either pure wino/bino-like
or where the lightest electroweakinos are of mostly higgsino nature.  The search
has discovery potential also when a light top squark and the LSP are nearly
degenerate in mass and the top squark decays to four fermions.  A more detailed
discussion of such models can be found in Ref.~\cite{Grober:2014aha}.  The
near-degeneracy in mass of the top squark and the LSP is typical of the
so-called ``co-annihilation region'', in which the LSP is the sole source of
dark matter~\cite{Coannihilation}.

In the models considered in this analysis, the visible decay products in the
SUSY signal have low momentum, which can be distinguished from SM processes when
a jet with large transverse momentum (\pt) from initial-state radiation (ISR)
leads to a large boost of the SUSY particle pair. This boost also enhances the
\ptmiss in the event. A similar search has previously been reported by the ATLAS
Collaboration~\cite{Aaboud:2017leg}. For the signal studied in this letter, SUSY
particles can decay leptonically, and the presence of low-\pt leptons can be
used to discriminate against otherwise dominant SM backgrounds, such as multijet
production through quantum chromodynamics (QCD) and \Zjets events with invisible
\Z boson decays.

The current strategy
is similar to that in the previous publication based on 8\TeV
data~\cite{Khachatryan:2015pot}, with the main difference being the deployment
of a new trigger selection that improves the sensitivity of the search in events
with two muons and low \ptmiss.
In addition, the selection has further been optimized for electroweakinos with a
compressed mass spectrum.
At least one jet is required in the final state; in the case of the signal, this jet must
arise from ISR, which provides the final-state particles with a boost in the
transverse plane, and
thereby the potential for moderate or large \ptmiss in the event.
Unlike the 8\TeV analysis, there is no upper limit on the number of jets in the event.
\section{CMS detector}\label{sec:cms}

The central feature of the CMS apparatus is a superconducting solenoid of
6\unit{m} internal diameter, providing a magnetic field of 3.8\unit{T}. Within
the solenoid volume are a silicon pixel and strip tracker, a lead tungstate
crystal electromagnetic calorimeter (ECAL), and a brass and scintillator hadron
calorimeter (HCAL), each composed of a barrel and two endcap sections. Forward
calorimeters extend the pseudorapidity ($\eta$) coverage provided by the barrel and
endcap detectors. Muons are detected in gas-ionization chambers embedded in the
steel flux-return yoke outside the solenoid.

Events of interest are selected using a two-tiered trigger system~\cite{Khachatryan:2016bia}.
The first level (L1), composed of custom hardware processors, uses information
from the calorimeters and muon detectors to select events at a rate of around
100\unit{kHz} within a time interval of less than 4\mus. The second level, known
as the high-level trigger (HLT), consists of a farm of processors running a
version of the full event reconstruction software optimized for fast processing,
and reduces the event rate to around 1\unit{kHz} before data storage.

A more detailed description of the CMS detector, together with a definition of
the coordinate system used and the relevant kinematic variables, can be found in
Ref.~\cite{Chatrchyan:2008zzk}.

\section{Data and simulated samples}
\label{sec:sample}

The data used in this search correspond to an integrated luminosity of
35.9\fbinv  of proton-proton (pp) collisions at a centre-of-mass energy of
13\TeV, recorded in 2016 using the CMS detector.  The data are selected using
two triggers: an inclusive \ptmiss trigger, which is used for signal regions
(SRs) with an offline \ptmiss cut $> 200$\GeV and an additional trigger which
requires two muons to lower the offline \ptmiss cut to $125$\GeV. Both the muon
\pt and the muon pair \pt have a trigger online cut of $\pt > 3$\GeV.  The
inclusive \ptmiss triggers correspond to an integrated luminosity of 35.9\fbinv,
whereas the events recorded with the dimuon+\ptmiss trigger correspond to
33.2\fbinv.

Simulated signal and major background processes, such as \ttbar, \PW+jets,
and \cPZ+jets are generated with the \MGvATNLO 2.2.2~\cite{Alwall:2014hca,Frederix:2012ps}
event generator at leading order (LO) precision in perturbative QCD using the
MLM merging scheme~\cite{Alwall:2007fs}. Additional
partons are modelled in these samples.
The diboson processes \PW\PW, \cPZ\cPZ, and $\PW\gamma$ are generated with the \MGvATNLO\!2.2.2 event generator at next-to-leading order (NLO) precision using the {FxFx} merging scheme~\cite{Frederix:2012ps},
while the \PW\Z process is generated at NLO with
{\POWHEG}~v2.0~\cite{Nason:2004rx,Frixione:2007vw,Alioli:2010xd,Melia:2011tj,Nason:2013ydw}.
Rare background processes (\eg $\ttbar\PW$, $\ttbar\Z$,
\PW\PW\PW, \cPZ\cPZ\cPZ, \PW\cPZ\cPZ, and \PW\PW\Z) are also generated at NLO precision
with \MGvATNLO\!2.2.2 (2.3.2.2
for~$\ttbar\cPZ$)~\cite{Alwall:2014hca,Frederix:2012ps}. The rare background
from single top quarks produced in association with a \PW{} boson is generated at
NLO precision with {\POWHEG}~v1.0~\cite{Re:2010bp}.
The NNPDF3.0~\cite{Ball:2014uwa} LO and NLO parton distribution functions (PDF) are used for the simulated samples generated
at LO and NLO. Showering, hadronization and the underlying event description
are carried out using the \PYTHIA~8.212 package~\cite{Sjostrand:2014zea}
with the CUETP8M1 underlying event tune~\cite{Skands:2014pea,CMS-PAS-GEN-14-001}.
A detailed simulation of the CMS detector is based on the \GEANTfour~\cite{Agostinelli:2002hh} package.
A fast detector simulation~\cite{bib-cms-fastsim-02} is used for the large
number of signal samples, corresponding to different SUSY particle masses.
The trigger, lepton identification, and \PQb tagging efficiencies are
corrected in the simulation through application of scale factors measured in
dedicated data samples~\cite{CMS-PAS-BTV-15-001}.
Corrections for the use of the fast detector simulation are also applied.

\begin{figure}[!hbtp]
\centering
\includegraphics[width=0.44\textwidth]{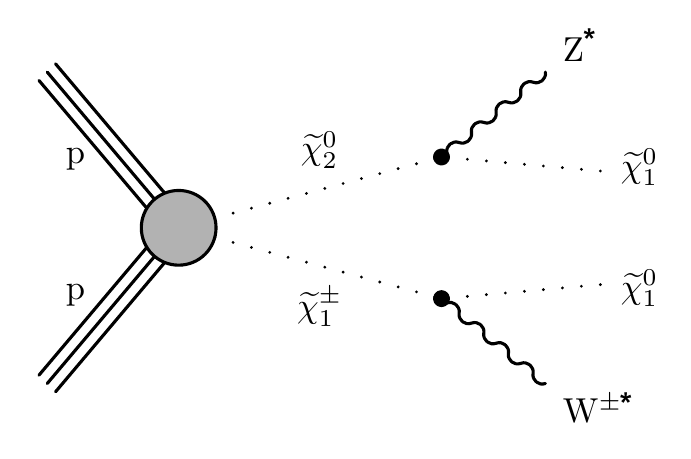}
\includegraphics[width=0.44\textwidth]{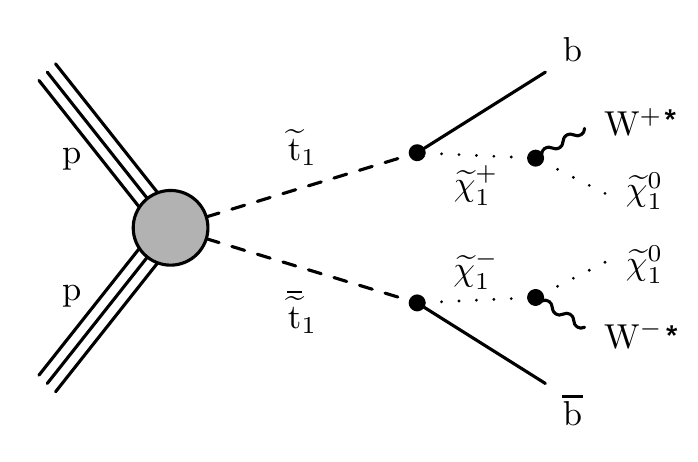}
\caption{Production and decay of an electroweakino pair (\cmsLeft) and of a chargino-mediated \sTop pair (\cmsRight).}
\label{Fig:Signals}
\end{figure}

For the signal, we consider the neutralino-chargino ($\chiz_2$-$\chipm_1$) pair
production where the mass degenerate $\chiz_2$ and $\chipm_1$ are assumed to decay to the LSP
via virtual \cPZ\ and \PW\ bosons. The decays of electroweakinos are carried out
using \PYTHIA,
assuming a constant matrix element.
The SM branching fractions
are assumed for the decays of the virtual \cPZ\ and \PW\ bosons.
The simulation of the $\chiz_2$ ($\chipm_1$) decay takes into account the Breit--Wigner shape
of the \cPZ\ (\PW) boson mass.
The production cross sections correspond to those of pure wino
production~\cite{Beenakker:1999xh,Fuks:2012qx,Fuks:2013vua} computed at NLO plus
next-to-leading-logarithmic (NLL) precision.
A second mass scan simulates a simplified model of \sTop-pair production, in
which a heavy chargino
mediates the decay of the \sTop into leptons and $\chiz_1$,
namely $\sTop\to \cPqb\chipm_1 \to \cPqb\PW^{*}\chiz_1$.
The mass of the $\chipm_1$ is set to $(m_{\sTop} + m_{\chiz_1})/2$, and the mass difference
between \sTop and $\chiz_1$ is set to be less than 80\GeV, thus b jets are
expected to have a \pt below 25\GeV.
Figure~\ref{Fig:Signals} shows diagrams for these two simplified models. We
denote the \cmsLeft diagram in Fig.~\ref{Fig:Signals} as TChi and the \cmsRight diagram
as T2tt. The masses are given with the model name, \ie TChi150/20
(T2tt150/20) denotes a $\chiz_2$-$\chipm_1$ ($\sTop$ pair) production, where the produced
particles have a mass of 150\GeV and a mass difference to the LSP of 20 GeV.

We interpret the results of this search in two variations of the
electroweakino model. While the model described above uses pure wino cross
sections with the $\chiz_2$ and $\chipm_1$ mass degenerate, these additional models
resemble a scenario where the electroweakinos are of higgsino nature. The
first of these higgsino simplified models features associated $\chiz_2$ and
$\chipm_1$ production and as such corresponds to the same diagram as the one
shown in  in Fig.~\ref{Fig:Signals} (\cmsLeft). The second higgsino model considers
associated $\chiz_2$-$\chiz_1$ production.
In both cases, the mass of the chargino is given as $m_{\chipm_1} = (m_{\chiz_2}
+ m_{\chiz_1})/2$, and the $\chiz_2$
decays via an off-shell \cPZ\ boson, and if applicable, the $\chipm_1$ decays via an off-shell \PW\ boson.
The simplified models do not include any spin correlations in the decays.
In the simplified higgsino model, this can lead to a different $\Mll$
distribution that we do not account for.

In addition to the electroweakino models, we interpret the results in a
phenomenological minimal supersymmetric model (pMSSM)~\cite{Djouadi:1998di}, in
which the higgsino ($\mu$), bino ($M_1$), and wino ($M_2$) mass parameters are
varied. There is only a small dependency on $\tan{\beta}$, which is set to 10.
All other mass parameters are assumed to be decoupled. To reduce the parameter
space to a two-dimensional grid, $M_2$ is set to $2 M_1$. This convention is
inspired by electroweakino mass unification at the grand unified theory scale.
Since the focus is on electroweak production only, the gluino mass parameter
$M_3$ is assumed to be decoupled. All trilinear couplings are discarded. In this
model, the higgsino mass parameter $\mu$ is varied between 100 and 200\GeV,
while $M_1$ varies between 300\GeV and 1\TeV. Events for this ``higgsino pMSSM''
are generated with \MGvATNLO\!\cite{Alwall:2011uj}. The NLO cross sections are
computed using Prospino 2~\cite{Beenakker:1996ed}.  Several additional
packages~\cite{Djouadi:2002ze,Muhlleitner:2003vg,Djouadi:1997yw,Djouadi:2006bz,Skands:2003cj}
are used to calculate mass spectra and particle decays.

\section{Object reconstruction}\label{section:objects}

The analysis makes use of the particle-flow (PF) algorithm~\cite{Sirunyan:2017ulk},
which reconstructs and identifies each individual particle through an optimized combination of information
from the various elements of the CMS detector.
The difficulties in reconstructing the event of interest,
because of the presence of the large average number of interactions per bunch crossing (pileup), are mitigated
by a primary vertex selection and other methods described below.
The reconstructed vertex with the largest value of summed physics-object $\pt^2$
is taken to be the primary $\Pp\Pp$ interaction vertex. The physics objects are
the jets, clustered using the jet finding
algorithm~\cite{Cacciari:2008gp,Cacciari:2011ma} with the tracks assigned to the
vertex as inputs, and the associated \ptmiss, taken as the
negative vector \pt sum of those jets.

The leading and subleading muon (electron) are required to satisfy $\pt>5\GeV$,
$\abs{\eta}<2.4$ ($2.5$).  A requirement of $\pt<30\GeV$ on the leptons is also
applied; this threshold is identified as the $\pt$ value below which the current
analysis is more sensitive in the compressed regions compared to other CMS
analyses.  To increase the sensitivity in the compressed mass regime, the lower
threshold on the $\pt$ of the subleading muon is set to 3.5\GeV in the
high-\ptmiss regions of the $\sTop$ search.

Muons are required to satisfy standard identification
criteria~\cite{Chatrchyan:2012xi},
and to be isolated within a cone in $\eta$--$\phi$ space of radius
$\Delta R = \sqrt{\smash[b]{(\Delta\eta)^2+(\Delta\phi)^2}}=0.3$: the $\pt$ sum of other charged particle tracks within the cone, $\text{Iso}_{\text{abs}}$,
is required to be less than 5\GeV.
In addition, the quantity $\text{Iso}_{\text{rel}}$, which is the ratio of
$\text{Iso}_{\text{abs}}$ and the \pt of the muon, is required to be less than 0.5.
Contamination from pileup within the isolation cone is subtracted using
techniques that utilize charged particle deposits within the cone itself~\cite{Chatrchyan:2012xi}.

Electrons from prompt decays are selected
using a multivariate discriminant based on the energy distribution in the shower and track quality variables.
The loose working point employed by the $\PH\to\cPZ\cPZ^{*}\to 4\ell$
analysis~\cite{CMS-HIG-16-041} is used for $\pt<10\GeV$, and a tighter one for $\pt>10\GeV$.
The same definition of isolation and the same isolation criteria are applied for electrons as used for muons.

To suppress nonprompt leptons, requirements on the three-dimensional impact
parameter~\cite{TRK-11-001} relative to
the primary vertex, $\mathrm{IP}_\text{3D}$, and its significance, $\mathrm{SIP}_\text{3D}$, are applied.
Leptons are required to have $\mathrm{IP}_\text{3D}<0.01$\unit{cm} and $\mathrm{SIP}_\text{3D}<2$
standard deviations (s.d.).

The combined efficiency for reconstruction, selection and isolation depends on
the $\pt$ of the lepton. The efficiencies are in the range 70\% (50\%) for muons
(electrons) at 5\GeV, up to 80\%\,(60\%) for muons\,(electrons) at 30\GeV.

Jets are clustered using the anti-\kt algorithm~\cite{Cacciari:2008gp} with a distance parameter
of 0.4~\cite{Chatrchyan:2011ds}, as implemented in the \FASTJET package~\cite{Cacciari:2011ma}.
The momentum of a jet, which is determined by the vectorial sum of all particle momenta in the jet,
is found from simulation to be within 5 to 10\% of the true momentum over the full \pt spectrum
and detector acceptance. An offset correction is applied to jet energies
to take into account the contribution from pileup~\cite{Cacciari:2007fd}.
Jet energy corrections are obtained from simulation, and confirmed through in situ
measurements of the energy balance in dijet and photon+jet events~\cite{Khachatryan:2016kdb}.
Jets are selected with $\pt>25\GeV$ and $\abs{\eta}<2.4$.
In the following, the transverse hadronic energy, $\HT$, is defined as the
scalar $\pt$ sum of the selected jets.

Jets arising from the hadronization of \PQb quarks are identified through the combined secondary
vertex (CSV) tagger~\cite{Sirunyan:2017ezt,Chatrchyan:2012jua},
which employs both secondary vertex and track-based information.
In this analysis, a loose working point corresponding to a \PQb tagging efficiency
of about 80\% is used with misidentification rates of 10\% and 40\% for light-quark or
gluon jets and for c quark jets, respectively~\cite{BTV-16-002}.

The \ptvecmiss is determined using the
PF-reconstructed objects. A variety of event filters are applied to remove
detector- and beam related noise~\cite{CMS-PAS-JME-16-004}.

\section{Event selection}\label{sec:evtReconstruction}
The analysis requires two oppositely charged leptons ($N_{\ell}=2$),
of either same ($\Pe\Pe$, $\mu\mu$) or different flavour ($\Pe\mu$),
and moderate \ptmiss in the final state,
together with at least one jet in the event.

The main backgrounds arise from events in which one of the leptons is not prompt
(mainly from \Wjets events), events from fully leptonic \ttbar decays ($\ttbar (2\ell)$),
and Drell--Yan (DY) processes with subsequent decays $\gamma/\cPZ^* \to \tau\tau \to \ell\ell\nu_{\ell}\nu_{\ell}\nu_{\tau}\nu_{\tau}$.
Smaller backgrounds are from $\cPqt\PW$ production (t\PW) and the diboson
processes $\PW\PW$ and $\cPZ\cPZ^*$, with $\cPZ^* \to \ell\ell$ and $\cPZ \to\nu\nu$ (VV).
Processes such as $\ttbar\PW$, $\ttbar\cPZ$, \PW\PW\PW, \cPZ\cPZ\cPZ,
\PW\cPZ\cPZ\, and~\PW\PW\cPZ\, as well as processes
including the Higgs boson have very small contributions, and are grouped
together as ``Rare''.
The following event selection shown in Table~\ref{tab:CUTS} includes a number of requirements designed to reduce these backgrounds:
\begin{itemize}
  \item $0.6<\ptmiss/\HT<1.4$: this criterion is effective
    in rejecting SM events comprised uniquely of jets produced through the strong
    interaction, referred to as QCD multijet events,
    while remaining efficient
    for events with ISR, as in the case of the signal. The bounds on the ratio $\ptmiss/\HT$ is determined
    from a study of a control region (CR) at low-\ptmiss and with dimuon mass close to that of the J/$\psi$ meson.
    This requirement rejects such events while leaving the signal unaffected.
  \item $\PQb$ jet event veto: requiring events where no jet is tagged as
    originating from $\PQb$ quarks
    significantly reduces the \ttbar background in which $\PQb$ jets originate
    from the decay of the top quarks.
    This requirement is applied to all jets with $\pt>25\GeV$ and uses the $\PQb$
    tagging selection criteria described in Section~\ref{section:objects}.
    The efficiency for a potential signal from \sTop decays is not affected significantly since in the compressed
    \sTop-LSP model, the $\PQb$ jets are expected to have small \pt and are
    therefore not tagged.
  \item $M(\tau\tau)<0$ or $M(\tau\tau)>160\GeV$:  this requirement on the
    estimate of the ditau mass is designed to reject the large background
    from $\cPZ\to\tau\tau$ decays, with the $\tau$ leptons decaying leptonically.  The quantity
    $M(\tau\tau)$~\cite{Han:2014kaa} is computed as follows: since the $\tau$ leptons from the decay of a $\cPZ$ boson have
    large \pt compared to their mass, the direction of the outgoing lepton
    is approximately the same as that of the $\tau$ lepton
    (\ie $\Delta R(\ell,\tau)\approx 0$). The magnitudes of the lepton momentum
    vectors are then rescaled so that the lepton
    pair balances the hadronic recoil.
    For $\cPZ\to\tau\tau$ events, this leads to a fairly good approximation of the original $\tau$ momenta.
    The invariant mass of the two $\tau$ leptons, $M(\tau\tau)$, is estimated by the invariant mass of the
    two scaled leptons. In some events, the estimate of the magnitude of the $\tau$ momentum results in a
    negative value when the flight direction is opposite to the direction of the lepton.
    In such cases, $M(\tau\tau)$ is set to its negative value.
  \item $\MT(\ell_{i},\ptmiss)<70\GeV$, for $i=1,2$: the transverse mass $\MT$ is
    defined as
    \begin{equation*}
      \MT(\ell, \ptmiss) = \sqrt{2 p_T^{\ell} \ptmiss \left(1 - \cos{\left[\Delta\phi\left(\ell,
      \ptmiss\right)\right]}\right)},
    \end{equation*}
    and $\ell_1$ and $\ell_2$ are the leading and subleading leptons,
    respectively.
    For the signal, the leading lepton is typically aligned with the boost direction
    of the LSP ($\Delta\phi(\ell,\ptmiss)\approx 0$). This requirement is effective in further suppressing
    the \ttbar background for the electroweakino search, but not for the \sTop
    search. It is therefore only applied in the electroweakino search.
  \item J$/\psi$, and $\Upsilon$ veto: to suppress background contributions from
    J$/\psi$, low-mass $\gamma^*$, and $\Upsilon$ decays, the dilepton invariant mass $M(\ell\ell)$ is required to satisfy $M(\ell\ell)>4\GeV$
    and to also lie outside the range $9<M(\ell\ell)<10.5\GeV$. This veto is
    only applied to same flavour lepton pairs.
  \item $\ptmiss>125\GeV$: to ensure high trigger efficiency,
    both the \ptmiss and the muon corrected $\ptmiss$, which is computed from the vectorial sum
    of the \ptmiss and the \pt of the muons selected in the event, is required to be larger than 125\GeV.
    The region $125\GeV < \ptmiss < 200$\GeV is only accessible by the dimuon
    trigger and therefore only dimuon pairs are considered. The region $\ptmiss
    > 200$\GeV includes also electrons.
  \item Trigger acceptance: in the online selection, the lepton pair is required to have a small boost of
    $\pt>3\GeV$, together with an upper bound on the dimuon invariant mass $M(\ell\ell)<60\GeV$,
    to limit the trigger rate. To remain fully efficient after offline reconstruction,
    an upper bound of 50\GeV on $M(\ell\ell)$ and a lower requirement
    on the dilepton transverse momentum $\pt (\ell\ell)>3\GeV$ are imposed.
  \item $\HT > 100\GeV$: this requirement suppresses backgrounds with low hadronic activity in the event.
\end{itemize}

\begin{table*}[!htbp]
  \topcaption{Common selection requirements for the signal regions.
    The subleading lepton \pt threshold is reduced to 3.5\GeV for muons in the
    high-\ptmiss, \sTop-like signal region.
  }
  \label{tab:CUTS}
  \centering
    \begin{tabular}{ll}
      Variable & SR selection criteria   \\  \hline
      $N_{\ell}$           & $2$ ($\mu\mu$, $\mu\Pe$, $\Pe\Pe$)           \\

      q($\ell_1$)q($\ell_2$)& $-1$             \\
      $\pt (\ell_{1})$, $\pt (\ell_{2})$     & $[5, 30]$\GeV \\
      $\pt (\mu_{2})$ for high-\ptmiss \sTop-like SR  & $[3.5, 30]$\GeV \\
      $|\eta_{\mu}|$& $<$2.4           \\
      $|\eta_{\Pe}|$& $<$2.5           \\
      $\mathrm{IP}_\text{3D}$ & $<$0.01\unit{cm}  \\
      $\mathrm{SIP}_\text{3D}$ & $<$2 \\
      $\text{Iso}_{\text{rel}}(\ell_{1,2})$ & $<$0.5 \\
      $\text{Iso}_{\text{abs}}(\ell_{1,2})$ & $<$5\GeV \\[\cmsTabSkip]
      $\pt (\text{jet})$      & $>$25\GeV \\
      $\abs{\eta} (\text{jet})$   & $<$2.4      \\
      $N_{\cPqb}$ ($\pt>$25\GeV, CSV) & $0$ \\[\cmsTabSkip]
      $M(\ell\ell)$ & $[4, 9]$ or $[10.5, 50]$\GeV (for $\mu\mu$ and $\Pe\Pe$)\\
      $\pt(\ell\ell)$       & $>$3\GeV \\
      $\ptmiss$       & $>$125\GeV (for $\mu\mu$) \\
                      & $>$200\GeV (for $\mu$e, ee) \\
      $\ptmiss$ (muon corrected) & $>$125\GeV (for $\mu\mu$) \\
                      & $>$200\GeV (for $\mu$e, ee) \\
      $\ptmiss/\HT$    & $[0.6, 1.4]$ \\
      $\HT$&$>$100\GeV\\
      $M(\tau\tau)$ & veto $[0,160]$\GeV \\
      $\MT(\ell_{i},\ptmiss), i=1,2 $ & $<$70\GeV (electroweakino selection only) \\
    \end{tabular}
\end{table*}

For the selected events, a set of SRs are defined, based on the dilepton
invariant mass and \ptmiss.
For events with leptons of same flavour and opposite charge, four SRs are defined in $M(\ell\ell)$ ranges
of 4--9, 10.5--20, 20--30, and 30--50\GeV. These SRs are intended for searches for $\chiz_2\to \cPZ^*\chiz_1$ events,
where $M(\ell\ell)$ is related to the mass difference between the two electroweakinos.
For events with leptons of different flavour and opposite charge, three SRs are defined in the
leading lepton \pt ranges of 5--12, 12--20, and 20--30\GeV. The definition of
the bins of the SRs can be found in Table~\ref{tab:srs}.

\begin{table*}[!htbp]
  \topcaption{Definition of bins in the two SRs. The lowest \ptmiss region
  includes only muon pairs, since it is only accessible by the dimuon trigger.}
  \label{tab:srs}
  \centering
    \begin{tabular}{ c  c {c}@{\hspace*{5pt}} c  c }
      \multicolumn{2}{ c }{Electroweakino search region}
        && \multicolumn{2}{c }{\sTop search region} \\
      $\ptmiss [\GeVns{}]$
        & $\Mll [\GeVns{}]$
        && $\ptmiss [\GeVns{}]$
        & $\pt^\text{lepton} [\GeVns{}]$ \\ \cline{1-2} \cline{4-5}
      \multirow{4}{*}{$[125, 200]$}
        & $[4, 9]$
        && \multirow{4}{*}{$[125, 200]$}
        & \multirow{4}{*}{\shortstack{$[5, 12]$ \\ $[12, 20]$ \\ $[20, 30]$}} \\
        & $[10.5, 20]$ &&& \\
        & $[20, 30]$ &&& \\
        & $[30, 50]$ &&& \\[\cmsTabSkip]
      \multirow{4}{*}{$[200, 250]$}
        & $[4, 9]$
        && \multirow{4}{*}{$[200, 300]$}
        & \multirow{4}{*}{\shortstack{$[5, 12]$ \\ $[12, 20]$ \\ $[20, 30]$}} \\
        & $[10.5, 20]$ &&& \\
        & $[20, 30]$ &&& \\
        & $[30, 50]$ &&& \\[\cmsTabSkip]
      \multirow{4}{*}{$>$250}
        & $[4, 9]$
        && \multirow{4}{*}{$>$300}
        & \multirow{4}{*}{\shortstack{$[5, 12]$ \\ $[12, 20]$ \\ $[20, 30]$}} \\
        & $[10.5, 20]$ &&& \\
        & $[20, 30]$ &&& \\
        & $[30, 50]$ &&& \\
    \end{tabular}
\end{table*}

To exploit the potential of the dimuon plus \ptmiss trigger, events are
separated according to the value of \ptmiss: in total three ranges are used for
the signal regions, namely $\ptmiss \in 125$--200, 200--300, and $>$300\GeV for
the \sTop search, and $\ptmiss \in 125$--200, 200--250, and $>$250\GeV for the
electroweakino search.  Since the low-\ptmiss region contains events accessible
only via the dimuon+\ptmiss trigger, only $\mu\mu$ pairs are considered. The
muons need to be of opposite charge. Conversely, in the high-\ptmiss regions,
both electron and muon flavours are considered. The electroweakino SRs are
populated by $\Pe\Pe$ and $\mu\mu$ pairs, where the leptons are oppositely
charged. For the \sTop SRs, $\Pe\mu$ pairs are
also considered. For the latter, the \pt threshold on the trailing lepton is
reduced to 3.5\GeV for muons in the high-\ptmiss region to gain sensitivity in
the search for \sTop signal.

The acceptance times efficiency for the signal model TChi150/20 (T2tt350/330) in
the electroweakino (stop) selection is between $3 \times 10^{-5}$\,($3 \times
10^{-5}$) and $7 \times 10^{-5}$\,($15 \times 10^{-5}$). The efficiency times
acceptance for muons is about 2 to 5 times higher than for electrons in the
electroweakino selection and about 1.5 to 3 times higher in the stop selection.

\section{Background estimation}
\label{sec:background}

Backgrounds with two prompt leptons are estimated using CRs chosen to be mostly
free from signal but when possible, with similar kinematic characteristics as
the events in the signal regions.  Different CRs are employed for each SM
process that contributes significantly to the signal region, \ie the \ttbar
dilepton background and the DY+jets background.  The normalisation of the
diboson background is cross checked in a validation region (VR).

For each background, the number of events in each SR is estimated
using the number of events observed in the corresponding CR,
and a transfer factor that is used to describe the expected ratio of events in the SR and CR
for the process in question.
The transfer factor for a specific process, $F_\text{process}$,
is determined from Monte Carlo (MC) simulation of the process through the ratio
$$F_\text{process} = \frac{N_{\text{MC process}}^\text{SR}}{N_{\text{MC process}}^\text{CR}}.$$

Since a CR typically contains contributions from other physics processes, they need to be subtracted
from the observed number of events in the CR, $N_\text{data}^\text{CR}$. These contributions, $N_{\text{MC other}}^\text{CR}$, are small compared to the
main process for which the CR is defined, and are thus estimated using MC simulation. The
estimate of the background from a specific physics process in the SR is then given by
\begin{equation*}
N_\text{process}^\text{SR} = \left( N_\text{data}^\text{CR} - N_{\text{MC other}}^\text{CR} \right) F_\text{process}.
\end{equation*}
Systematic uncertainties in the value of $F_\text{process}$ are included when determining the full
uncertainty in $N_\text{process}^\text{SR}$.  The total background in the SR is given as the sum of
the backgrounds expected from each process.

The different CRs are split into two \ptmiss bins: The low \ptmiss bin with
\ptmiss between 125 and 200\GeV is used to constrain the SRs with the same
\ptmiss range, while the high \ptmiss bin with \ptmiss $>$200\GeV is used to constrain all SRs
with \ptmiss above 200\GeV. The shapes for $\Mll$ and the lepton \pt are taken
directly from simulation.
A summary of all CRs for prompt lepton backgrounds is given in Table~\ref{tab:CR_cut_summary}.
For the diboson background, a validation region enriched in VV (mainly $\PW\PW$
events) is added.
This region is used to establish how well the simulation agrees with data in
order to validate the uncertainty assigned to the diboson simulation. About half
of the events in this region stem from VV.
\begin{table*}[!hbtp]
\centering
\topcaption{Summary of changes in selection criteria relative to
Table~\ref{tab:CUTS} for CRs and the VV validation region (VR).\label{tab:CR_cut_summary}}
\resizebox{1.0\textwidth}{!}{
\begin{tabular}{| c | c | c |}
 \multicolumn{1}{c}{DY CR} &   \multicolumn{1}{c}{\ttbar($2\ell$) CR} &  \multicolumn{1}{c}{VV VR} \\\hline
 \multicolumn{3}{|c|}{No upper requirement on \pt($\ell$) } \\
 \multicolumn{3}{|c|}{$\text{Iso}_\text{rel}<0.1$ as an or condition with the SR isolation } \\ \hline
 \multirow{4}{*}{\shortstack[c]{
  $0< M(\tau\tau) <160\GeV$ \\
  $\mathrm{IP}_\text{3D}<0.0175\unit{cm}$, $\mathrm{SIP}_\text{3D}<2.5$ s.d. \\
  $\pt(\ell_1)> 20\GeV$, or $\mathrm{IP}_\text{3D}>0.01\unit{cm}$, or $\mathrm{SIP}_\text{3D}>2$ s.d. \\
  \MT as for electroweakino SR
 }}
 & \multirow{4}{*}{\shortstack[c]{
   No requirements on \MT \\
   At least one b-tagged jet \\
   with $\pt > 40\GeV$
 }}
 & \multirow{4}{*}{\shortstack[c]{
  $\pt(\ell_1)> 20\GeV$ \\
  $|\text{same flavour}\,\Mll - M(\cPZ) | > 10\GeV$ \\
  $\MT > 90\GeV$
 }} \\ && \\ && \\ && \\
 \hline
\end{tabular}
}
\end{table*}

\subsection{The DY+jets control region}
The main difference between the CR for the DY+jets background and the SR
lies in the requirement imposed on the $M_{\tau\tau}$ variable; the CR consists of events that are
vetoed in the SR selection, namely those events with $M_{\tau\tau}$ in the range 0--160\GeV.
To increase the efficiency for leptons from $\tau$ decays, the
impact parameter requirements are relaxed to $\mathrm{IP}_\text{3D}<0.0175\unit{cm}$ and
$\mathrm{SIP}_\text{3D}<2.5$ s.d.
The variation of the scale factors applied to simulation by changing the cuts
on $\mathrm{IP}_\text{3D}$ and $\mathrm{SIP}_\text{3D}$ was found to be negligible.
In addition, the $30\GeV$ upper bound on the lepton \pt is removed, and the region
with lepton $\pt<20\GeV$, $\mathrm{IP}_\text{3D}<0.01\unit{cm}$, and $\mathrm{SIP}_\text{3D}<2$ is also
removed to reduce the presence of potential signal.
The distributions in kinematic quantities of these events,
including the variables used to define the signal regions, $\Mll$ and the leading lepton \pt,
are well described in simulation.
The event yields estimated from simulation and the observed event yields are listed in
Table~\ref{tab:CR_event_yields_DY}.

\begin{table*}[!hbtp]
\centering
\topcaption{Data and simulation yields for the DY and \ttbar($2\ell$) CRs,
  corresponding to integrated luminosities of 35.9\fbinv (high-\ptmiss region)
  and 33.2\fbinv (low-\ptmiss region). The SR scale factors are derived by
  subtracting the other processes from the observed data count, and dividing
  this number by the expected event yields from simulation for the process in
  question. The uncertainties are statistical only.}
\begin{tabular}{ l {c}@{\hspace*{5pt}} c  c {c}@{\hspace*{5pt}} c  c   }
  && \multicolumn{2}{c}{ DY CR }  &&  \multicolumn{2}{c}{ \ttbar($2\ell$) CR }  \\
  \multicolumn{1}{l}{\ptmiss}                       &&  \multicolumn{1}{c}{125--200\GeV}       &  $>$200\GeV      && \multicolumn{1}{c}{125--200\GeV}         &  $>$200\GeV    \\\cline{3-4}\cline{6-7}
 DY+jets or \ttbar             &&  70.1 $\pm$ 5.1     &  64.5 $\pm$ 3.3    && 1053.7 $\pm$ 9.4     &  535.7 $\pm$ 7.1  \\
 All SM processes              &&  82.6 $\pm$ 5.5     &  75.2 $\pm$ 3.6    && 1170.0 $\pm$ 11.0    &  710.4 $\pm$ 11.1 \\
 Data                          &&   84                &  75                && 1157                 &  680  \\
 SR scale factor               && 1.02 $\pm$ 0.13     &  0.99 $\pm$ 0.13   && 0.99 $\pm$ 0.03      &  0.94 $\pm$ 0.05    \\
\end{tabular}

\label{tab:CR_event_yields_DY}
\end{table*}

\subsection{The \texorpdfstring{\ttbar ($2\ell$)}{tt (2ll)} control region}

To obtain a sample enriched in \ttbar events, at least one jet is required to be
identified as originating from \PQb quarks.  To reduce potential signal
contamination, the leading \PQb-tagged jet is required to satisfy $\pt>40\GeV$.
To increase the number of events in the CR, while still avoiding potentially
large signal contamination, the upper bound on the lepton \pt is also removed.
The event yields estimated from simulation and the observed event yields are
also shown in Table~\ref{tab:CR_event_yields_DY}.

\subsection{Nonprompt background}

The background from nonprompt or misidentified leptons
is evaluated using a ``tight-to-loose'' method. Events where at least one lepton fails the
tight identification and isolation criteria but passes a looser selection define the ``application region''.
Events in this region are weighted by a transfer factor based on the probability that
nonprompt leptons passing the loose requirements also satisfy the tight ones.
The resulting estimate is corrected for the presence of prompt leptons in the application region.

The probability for nonprompt or misidentified leptons to pass the tight selection criteria is referred to as the
misidentification probability, which is determined as a function of lepton \pt and $\eta$.
This probability is measured using a dedicated
data sample, the ``measurement region'' (MR), which is enriched in the background from
SM events containing only jets produced via strong interaction, referred to as QCD multijet events.
This method has been used in several multilepton analyses at CMS and is described in more detail
in Ref.~\cite{Chatrchyan:2011wba}. The MR is defined through the presence of one loose lepton,
obtained by relaxing the isolation and impact parameter requirements, and
through a jet with
$\pt>30\GeV$, separated from the lepton by $\Delta R>0.7$. For muons, events are
selected through
prescaled single-lepton triggers with no isolation requirements. For electrons, a mixture of prescaled
jet triggers is used. The method includes a correction for the presence of prompt leptons in
the MR, mostly due to $\PW$ and $\Z$ boson production in association with jets.
The probability for prompt leptons to pass the tight selection criteria is taken from simulation
and is corrected with a data-to-simulation scale factor extracted from data enriched in
$\Z\to\ell\ell$ decays.

In this analysis, the misidentification probability measured in QCD multijet events is applied to loosely
identified leptons in events that are dominated by \Wjets and \ttbar production.
The latter can have both a different composition in terms of the flavour of the jets that give rise to the
nonprompt leptons, as well as different kinematic properties, potentially resulting in a different
effective misidentification probability. These effects are studied by comparing
the misidentification probabilities measured
in simulated events of these two processes in the kinematic regions
probed by this analysis.
A closure test is then performed by applying the misidentification probability measured in the QCD
simulated multijet events to a sample of \Wjets events. The yield of events passing the tight identification
criteria is compared with the estimate obtained by applying the
misidentification probability to events in the
application region.
The method is found to be consistent within a level of $<$40\%; this value is used
as a systematic uncertainty in the estimate of the normalization of the
reducible background.

\begin{figure}[!htb]
        \centering
\includegraphics[width=\cmsFigWidthTwo]{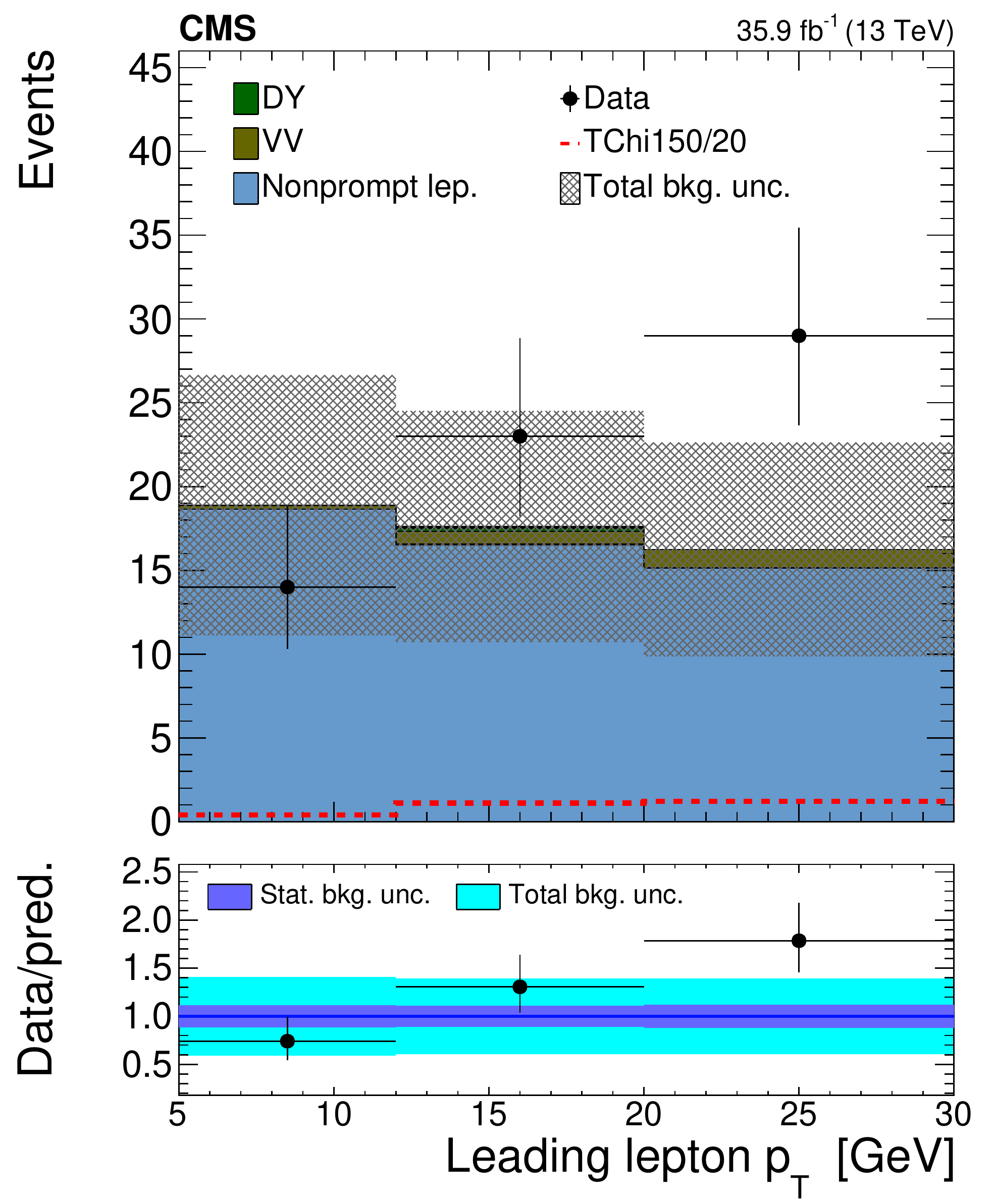}
        \caption{Same-sign CR for \sTop selection and $\ptmiss>200\GeV$.
        The distribution of the leading lepton $\pt$ is used as input to the final signal extraction.
        A signal from neutralino-chargino ($\chiz_2$-$\chipm_1$) production
      is superimposed.}
        \label{fig:SSCR}
\end{figure}

To further constrain the contribution of the nonprompt lepton background in the
SR, a dedicated CR consisting of same-sign (SS) leptons is defined.  Requiring
the two lepton candidates to have the same sign increases significantly the
probability that at least one of the two is a nonprompt or misidentified lepton.
The SS CR is defined using the \sTop selection in the $\ptmiss>200\GeV$ region,
where the opposite charge requirement of the two leptons is modified to
same-sign.  In the SS CR, the prediction of the nonprompt lepton background is
derived from the ``tight-to-loose" method and agrees with the data.
Figure~\ref{fig:SSCR} shows the leading lepton \pt distribution in the SS CR. It
also shows the near absence of a signal. The distribution of the leading lepton
\pt is used as input to the final fit that performs the signal extraction, as
its constraining power is significant, given the significant uncertainty on the
measured misidentification probability.

\section{Systematic uncertainties}
\label{sec:systematics}

This section summarizes the systematic uncertainties in the estimate of the background
from the various SM processes. For each source of systematic uncertainty, we
present
both the effect on the corresponding specific background and the overall effect on
the total background predictions are listed in Table~\ref{tab:SYSTEMATICS_1}.

\begin{table*}[htb]
\topcaption{Relative uncertainties in the final total background predictions for
  each individual systematic source of uncertainty.}
  \label{tab:SYSTEMATICS_1}
  \centering
  \begin{tabular}{l c  }
 Systematic source of uncertainty     & Typical uncertainty (\%)   \\ \hline
 VV background normalization          & 3--25 \\
 Nonprompt lepton background normalization & 4--20 \\
 DY+jets background normalization    & 4--20 \\
 \ttbar background normalization      & 2--8 \\
 Rare background normalization       & 1--3  \\
 Jet energy scale                     & 2--12  \\
 b tagging                            & 2--6  \\
 Pileup                               & 1--5 \\
 Lepton selection                     & 1--4 \\
 Integrated luminosity                & 2.5  \\
 Trigger                              & 1--2  \\
 \ttbar modelling                      & $<$1  \\
\end{tabular}
\end{table*}

The uncertainty in the predicted nonprompt lepton background contains a
statistical component due to the statistical uncertainty in the application
region event yield, it ranges from 10\% to 50\%. When applied in the SR, the
uncertainty is 4\% to 20\%.  Another source of statistical uncertainty arises
from limited statistics in data and simulation in the DY+jets and \ttbar
($2\ell$) CRs. The effect on the predicted yields in the SR, obtained using the
transfer factor described in Section~\ref{sec:background}, is approximately 13\%
for the DY+jets background and 3\% for the \ttbar background.

For the \ttbar background, we have considered a set of systematic uncertainties arising from
the modelling of the kinematic distributions in the simulation of this process.
The spin correlation of the top quarks has been varied by 20\%, based on the
ATLAS and CMS~\cite{Aad:2014mfk,Khachatryan:2016xws} measurements and a
comparison between different generators (\MGvATNLO versus \POWHEG).
The helicity amplitudes of the \PW\ boson in top quark decays have been varied by 5\%.
A top quark \pt modelling uncertainty has also been derived by reweighting the
simulated \ttbar events based on the number of ISR jets ($N_\text{jets}^{\mathrm{ISR}}$),
so as to make the jet multiplicity agree with data. The reweighting factors
range from
0.92 to 0.51 for $N_\text{jets}^{\mathrm{ISR}}$ between 1 and 6. The systematic uncertainty
in these reweighting factors is taken to be equal to one half of the deviation of the
factor from unity. The combined effect of this set of \ttbar modelling uncertainties on
the total number of predicted \ttbar background events is found to be in the range 3--5\%.

For the DY+jets background, the uncertainty in the resolution of the \pt of the system recoiling against
the two leptons is obtained from data dominated by $\cPZ\to\mu\mu$ events.
The uncertainty affects the DY estimate, which uses the efficiency of the requirements on $M_{\tau \tau}$
from simulation. The effect on the estimated yields of DY+jets is found to be negligible ($<$1\%).

As presented in Section~\ref{sec:background}, the method used to estimate the background from
nonprompt and misidentified leptons leads to a 40\% uncertainty on the
normalization. In the global fit this uncertainty is reduced to 25\%.

A 50\% uncertainty is assigned for the diboson background normalization, which is checked
in the dedicated region described in Section~\ref{sec:background}. In this region, which is
enriched in \PW\PW\ events with similar kinematic properties as the events in the SR,
the simulation is found to agree, within the given uncertainty, with the data.

A conservative 100\% uncertainty is assigned to the very small rare backgrounds that are
dominated by the \cPqt\PW\ process.

The experimental uncertainties related to \cPqb\ tagging, trigger, lepton reconstruction,
identification, and isolation criteria have been propagated and their effect on the final results
ranges from 2\% up to 12\%.
The jet energy scale corrections (JEC) are applied to match jet energies
measured in data and simulation. The JEC are affected by an intrinsic uncertainty, which affects all
simulated background, leading to typically 2--12\% uncertainties in the final predictions.

An uncertainty of 2.5\% is assigned to the integrated luminosity measured by CMS
for the 2016 data taking period~\cite{CMS-PAS-LUM-17-001}. This affects the estimate of the rare SM backgrounds
that rely on the measured data luminosity.

Finally, the uncertainty related to pileup has been estimated by varying the minimum-bias
cross section by $\pm$5\% and reweighting the pileup distribution accordingly.
The systematic uncertainty is found to be in the range 1--5\%.

As the signal yields are from simulation, additional systematic uncertainties
are applied in two categories. One arises from the systematic
uncertainty in the inclusive NLO+NLL~\cite{Beenakker:1999xh,Fuks:2012qx,Fuks:2013vua} cross section
used for the normalization, determined by varying the renormalization and factorization scales
and the PDF. The dependence on these QCD scales
yields a total uncertainty of 3\%.
The other category arises from the uncertainty in the product of
the signal acceptance and efficiency.

It is important to properly model the ISR that leads to the boost of the
produced SUSY particles in the transverse plane.
In particular, for the electroweakino benchmark, the modelling of the ISR with \MGvATNLO
affects the total transverse momentum $\pt^{\mathrm{ISR}}$ of the system of SUSY particles,
which can be improved by reweighting $\pt^{\mathrm{ISR}}$ in the simulated signal events.
This reweighting is based on $\pt$ studies of events containing a \cPZ\
boson~\cite{Chatrchyan:2013xna}, in which the factors range between 1.18 at
$\pt^{\mathrm{ISR}}$ of 125\GeV,
and 0.78 for $\pt^{\mathrm{ISR}}>600\GeV$. The deviation from 1.0 is taken as the systematic uncertainty
of the reweighting procedure.
For the \sTop benchmark to improve the modelling of the multiplicity of additional jets
from ISR, the events are reweighted based on the $N_\text{jets}^{\mathrm{ISR}}$,
using the same corrections used for the top background as described earlier in
this section.
The typical uncertainties on the final results from the ISR modelling are found to be in the range 2--7\%.

We account for differences observed in \ptmiss reconstruction effects in full and fast simulation used for signal.
The uncertainties vary between 3 and 5\%. The uncertainties related to potential differences
in \PQb tagging between the full and fast simulation and in the JEC vary in the range 1--2\%.

These uncertainties, together with those related to the predicted backgrounds described in Section~\ref{sec:background},
are included as log-normal distributed nuisance parameters in the likelihood approach.

\section{Results}

\begin{figure*}[h!t]
\centering
\includegraphics[width=0.32\textwidth]{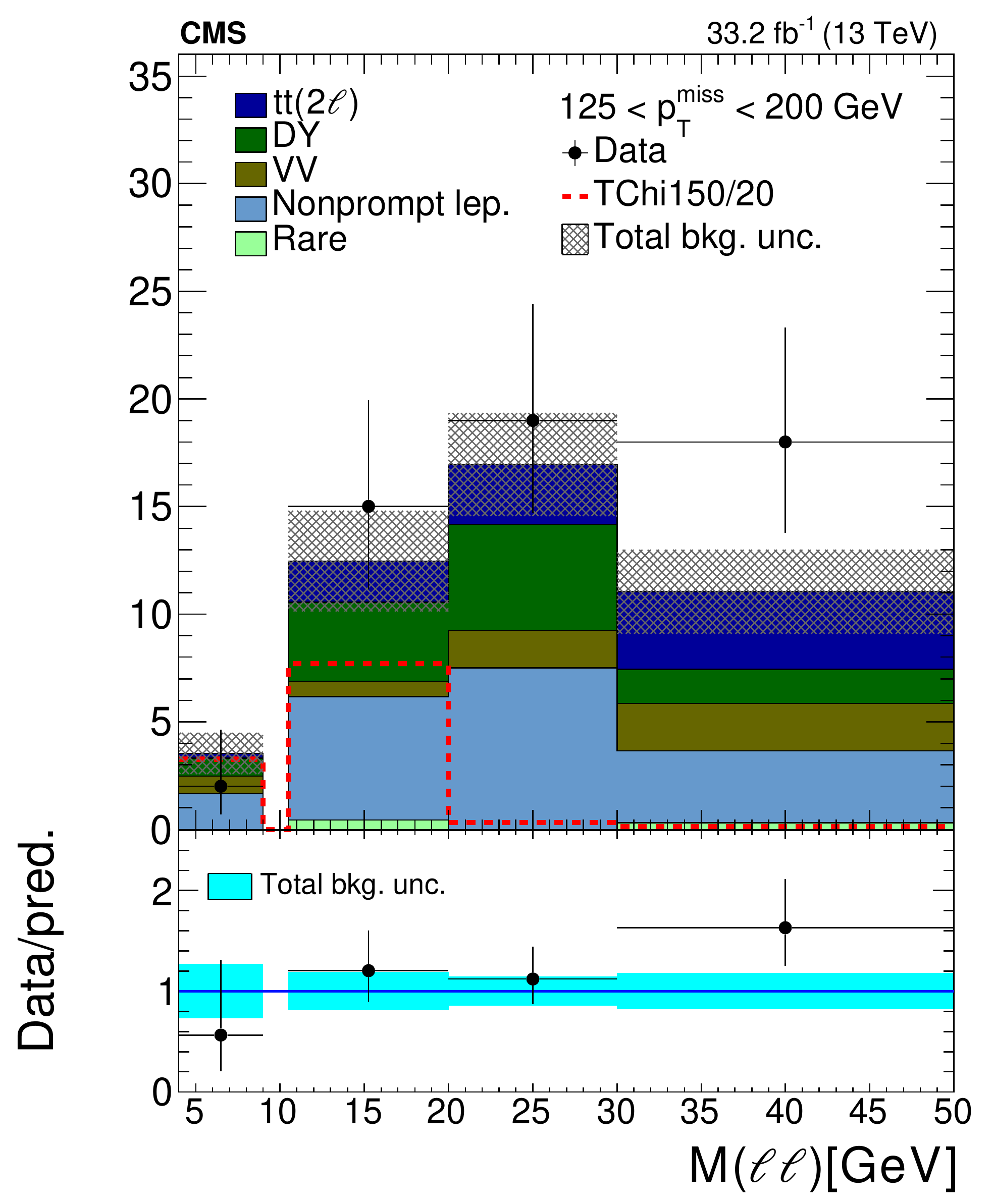}
\includegraphics[width=0.32\textwidth]{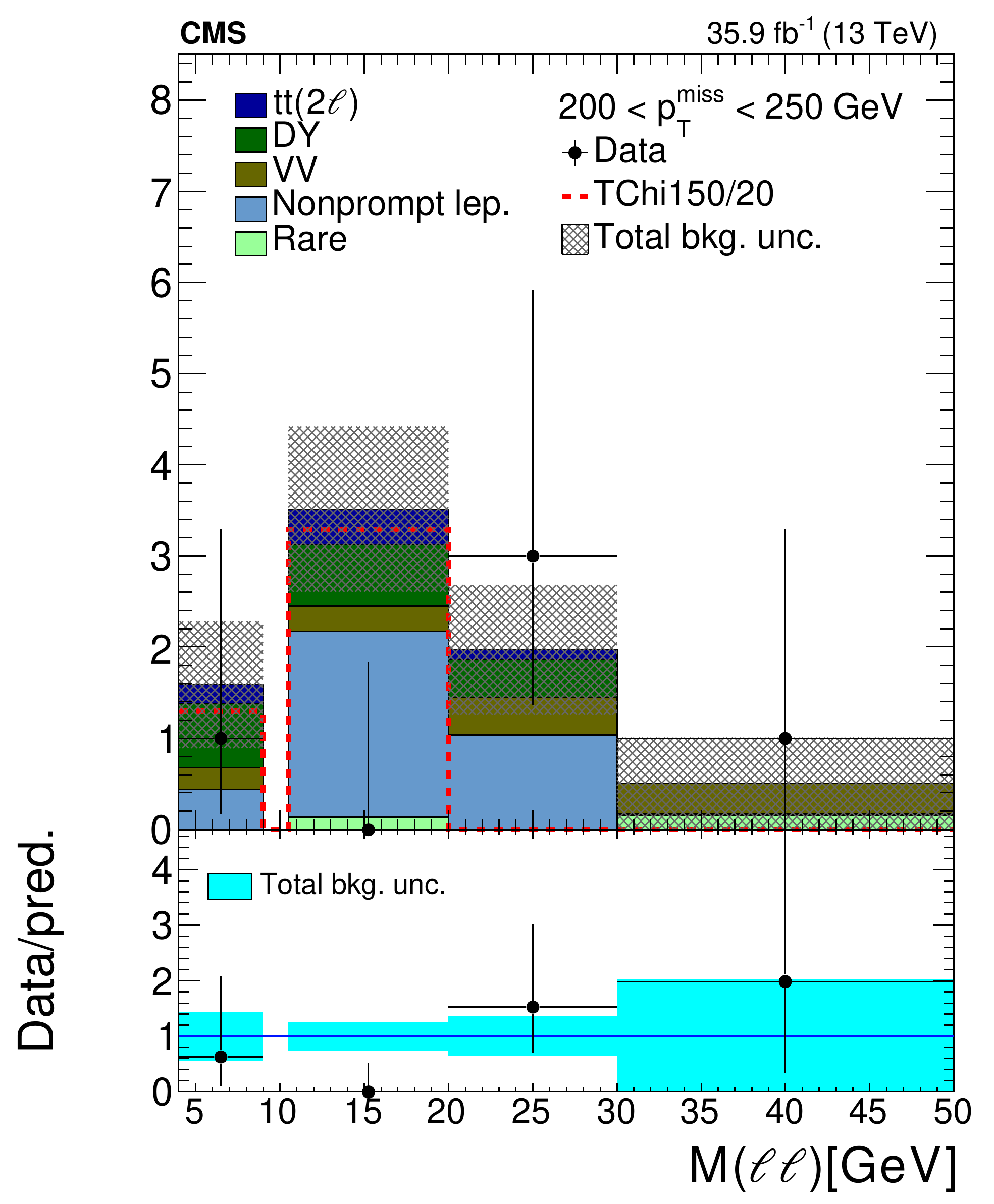}
\includegraphics[width=0.32\textwidth]{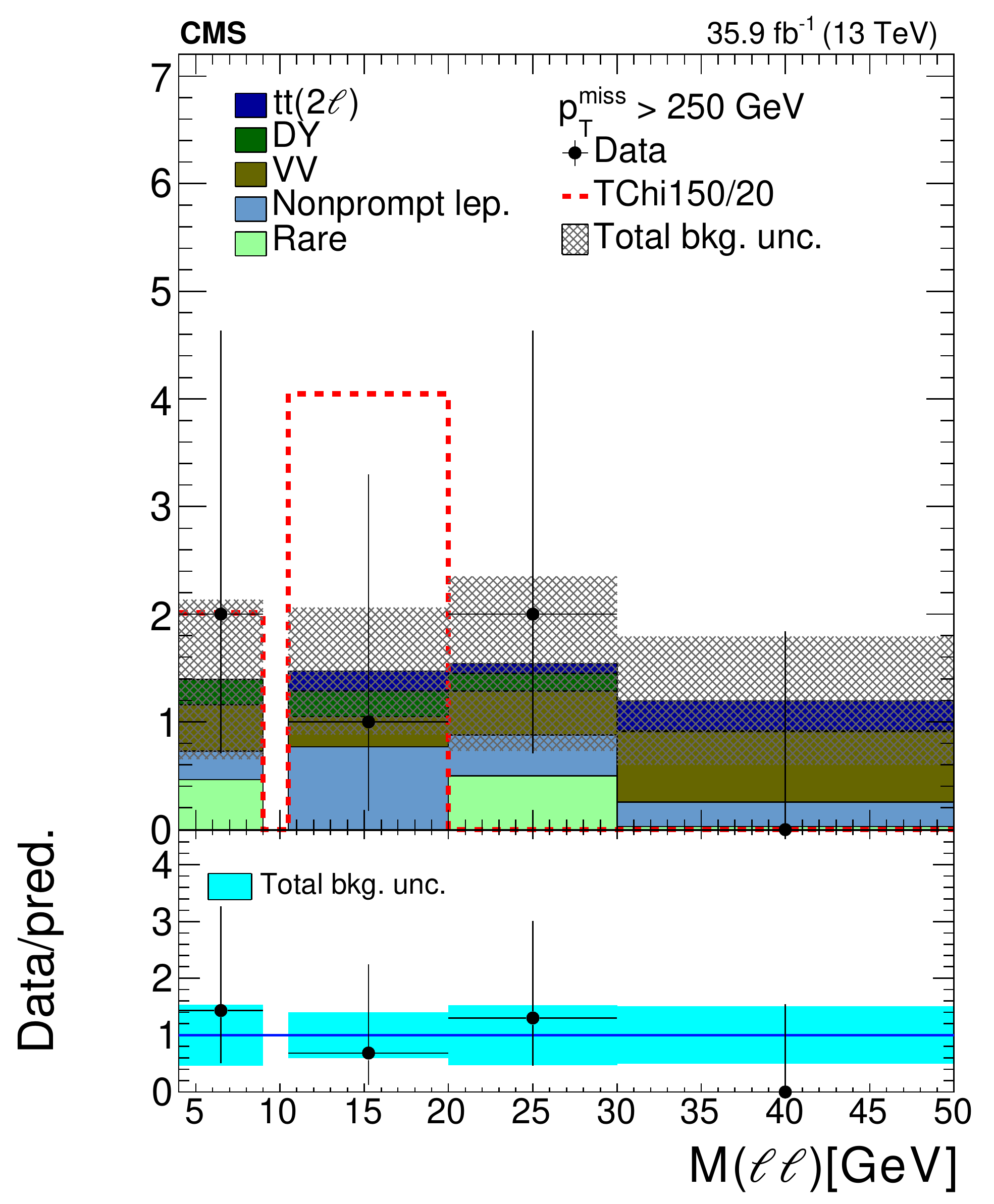}
\caption{Left: electroweakino search regions in bins of $M(\ell\ell)$ for $125<\ptmiss<200\GeV$ (muon only channel) for 33.2\fbinv;
middle: $200<\ptmiss<250\GeV$ (muon and electron channel) for 35.9\fbinv;
right: $\ptmiss>250\GeV$ (muon and electron channel) for 35.9\fbinv.
        A signal from neutralino-chargino ($\chiz_2$-$\chipm_1$) production
      is superimposed. The gap between 9 and 10.5\GeV corresponds to the
      $\Upsilon$ veto.}
\label{Fig:SR125_blind-ewkino}
\end{figure*}

\begin{figure*}[h!t]
\centering
\includegraphics[width=0.32\textwidth]{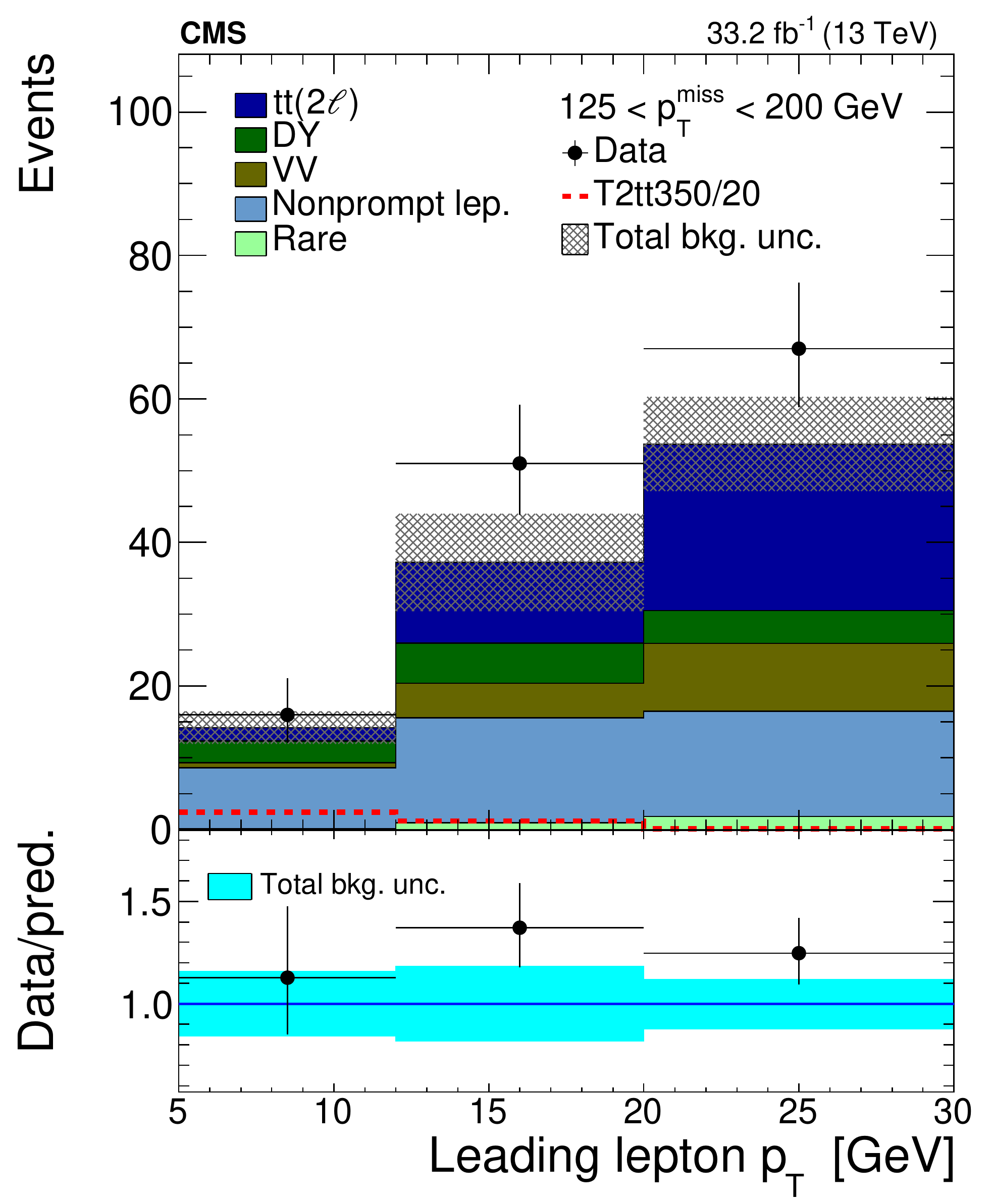}
\includegraphics[width=0.32\textwidth]{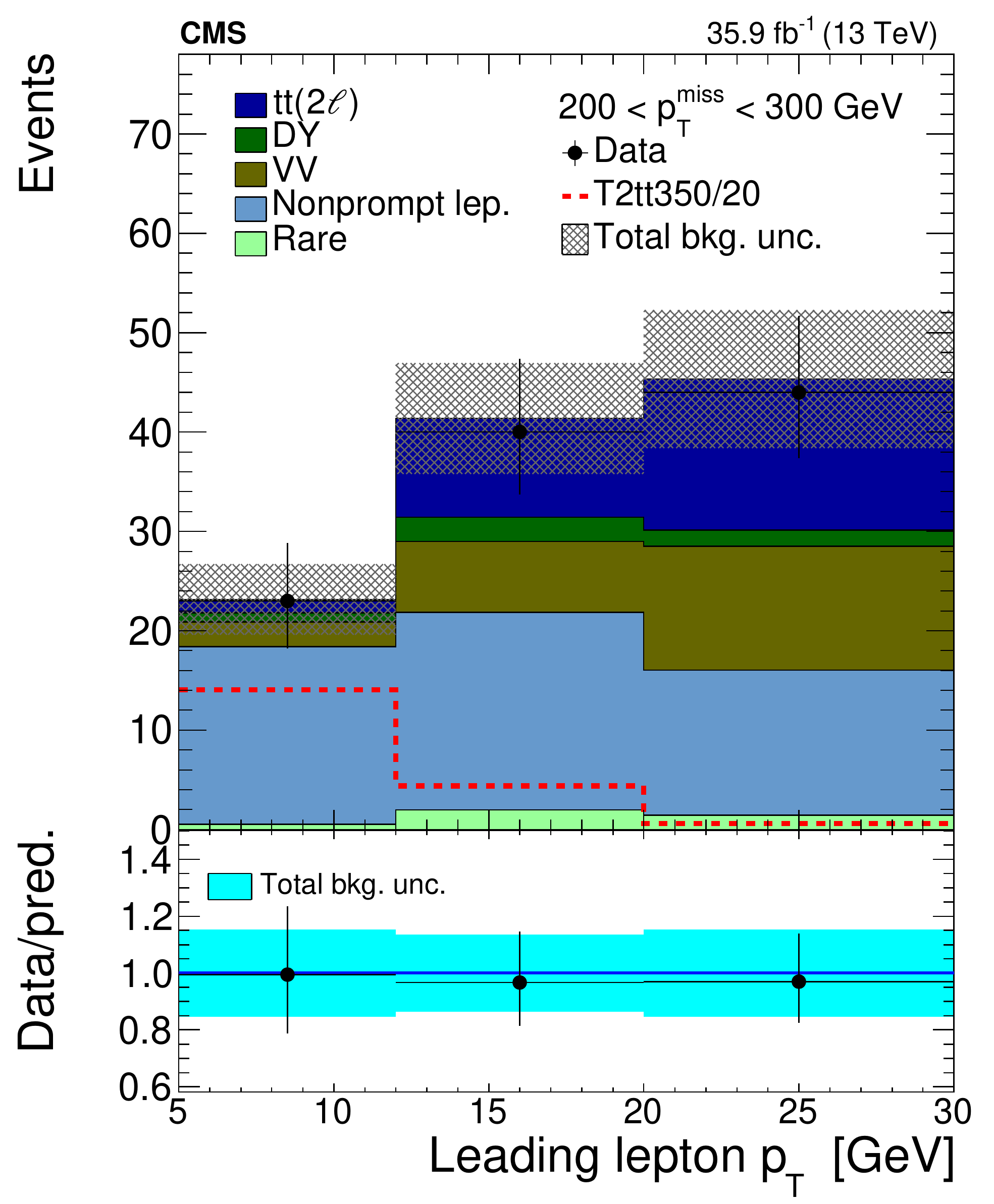}
\includegraphics[width=0.32\textwidth]{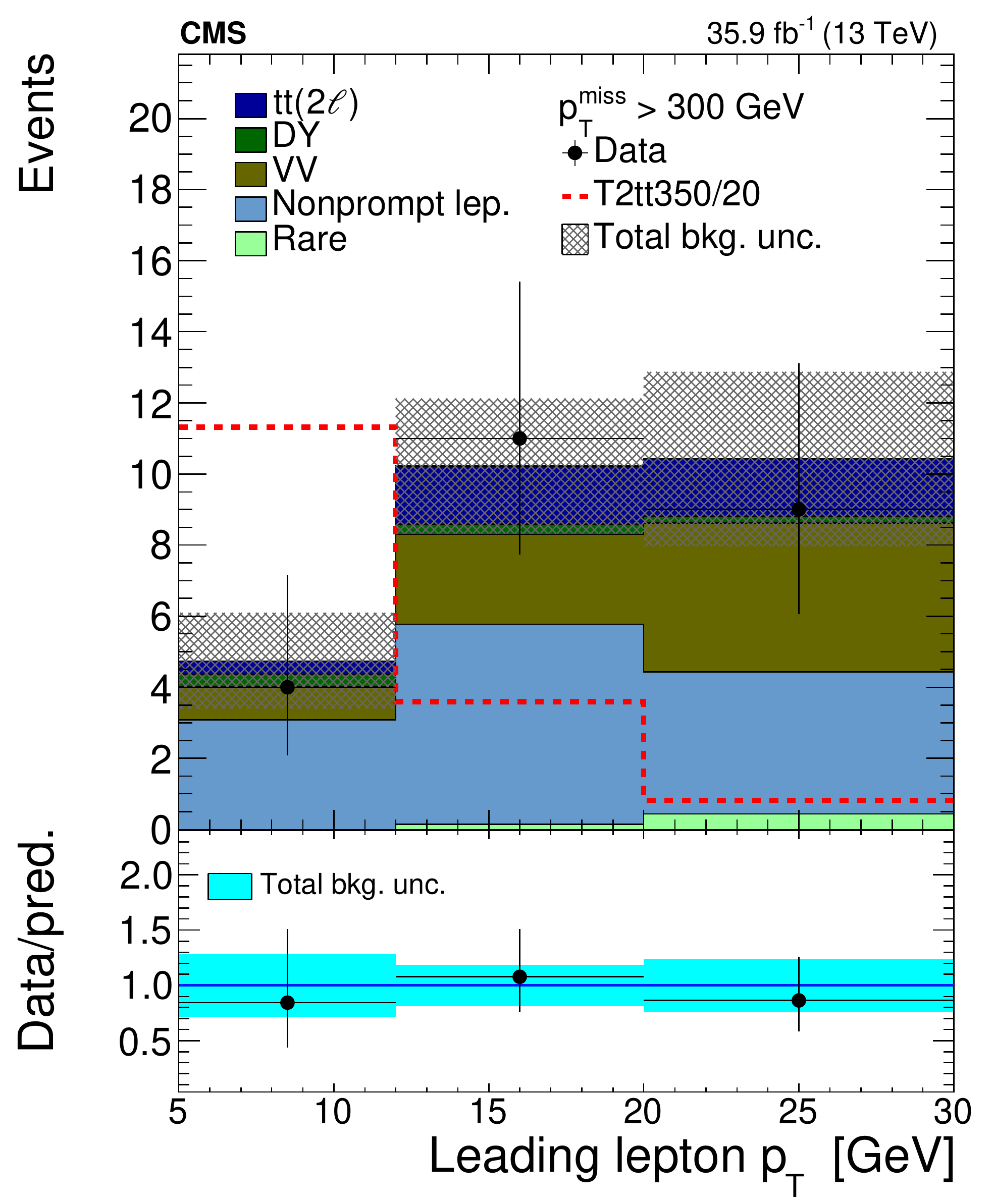}
\caption{Left: $\PSQt$ search regions in bins of leading lepton \pt for $125<\ptmiss<200\GeV$ (muon only channel) for 33.2\fbinv;
middle: $200<\ptmiss<300\GeV$ (muon and electron channel) for 35.9\fbinv;
right: $\ptmiss>300\GeV$ (muon and electron channel) for 35.9\fbinv.
        A signal from $\PSQt$ pair production
      is superimposed.}
\label{Fig:SR200_blind}
\end{figure*}

The estimated yields of the SM background processes and the data observed in the SRs
are shown in Figs.~\ref{Fig:SR125_blind-ewkino} and \ref{Fig:SR200_blind}.
No significant excess has been observed.
The estimates in the SR bins are extracted from a maximum likelihood fit of the data
using the expected yields described in Section~\ref{sec:background}, namely the DY+jets,
\ttbar ($2\ell$), and SS CRs.
Log-normal distributions for nuisance parameters are used to describe the systematic uncertainties of
Section~\ref{sec:systematics}. The uncertainties in the predicted
yields quoted in the following are those determined from the fit.

The predicted yields along with the data are also summarized in
Tables~\ref{tab:PREDh_SR_1} and~\ref{tab:PREDh_SR_2} for each bin of the SR.The
total uncertainty in the yield for each SM process includes the systematic and
statistical uncertainties described in Section~\ref{sec:systematics}, added in
quadrature. The largest deviation from the SM expectation is seen in a bin of
the electroweakino search region. The bin with $\ptmiss \in [200, 250]$\GeV and
$\Mll \in [10.5, 20]$\GeV has $3.5 \pm 0.9$ expected events but 0 observed. The
smaller number of events observed in this bin drives the observed exclusion to
higher values than expected, as can be seen in the next section. Overall, there
is good agreement between expectation and observation.

\begin{table*}[h!t]
\centering
\topcaption{The number of events observed in
the data and the result of the fit of the backgrounds to the data in the
electroweakino search regions.
The uncertainty indicated is determined from the fit to the 33.2 and 35.9\fbinv integrated
luminosities. Values for the \Mll ranges are in \GeV. Rare background event yields are omitted when they do not contribute to the SR bin.\label{tab:PREDh_SR_1}}
\cmsTabResize{
\begin{tabular}{l c c c c  }
                     &  \multicolumn{4}{c}{ $125<\ptmiss<200$\GeV } \\
                     &  $4<\Mll<9$ & $10.5<\Mll<20$ & $20<\Mll<30$  & $30<\Mll<50$   \\ \hline
 $\ttbar(2\ell)$  & 0.23 $\pm$ 0.16 & 1.9 $\pm$ 0.52 & 2.80 $\pm$ 0.65 & 3.60 $\pm$ 0.75 \\
 DY+jets  & 0.83 $\pm$ 0.63 & 3.7 $\pm$ 1.5 & 4.9 $\pm$ 1.5 & 1.60 $\pm$ 0.99 \\
 VV  & 0.82 $\pm$ 0.48 & 0.71 $\pm$ 0.65 & 1.7 $\pm$ 1.0 & 2.2 $\pm$ 1.2 \\
 Nonprompt lepton  & 1.7 $\pm$ 0.7 & 5.7 $\pm$ 1.5 & 7.5 $\pm$ 1.7 & 3.3 $\pm$ 1.1 \\
 Rare  & \NA & $0.46^{+0.64}_{-0.45}$ & \NA &  $0.33^{+0.49}_{-0.32}$ \\
 \textbf{Total SM prediction}  & 3.5 $\pm$ 1.0 & 12.0 $\pm$ 2.3 & 17.0 $\pm$ 2.4 & 11.0 $\pm$ 2.0 \\[\cmsTabSkip]
 Data  & 2 & 15 & 19 & 18 \\  \hline
\\
                     &  \multicolumn{4}{c}{ $200<\ptmiss<250$\GeV} \\
                     &  $4<\Mll<9$ & $10.5<\Mll<20$ & $20<\Mll<30$  & $30<\Mll<50$  \\ \hline
 $\ttbar(2\ell)$  & 0.21 $\pm$ 0.17 & 0.38 $\pm$ 0.18 & $0.11^{+0.11}_{-0.10}$ & \NA \\
 DY+jets  & 0.69 $\pm$ 0.62 & 0.67 $\pm$ 0.32 & 0.42 $\pm$ 0.27 & \NA \\
 VV  & $0.26^{+0.28}_{-0.25}$ & $0.29^{+0.32}_{-0.28}$ & 0.42 $\pm$ 0.33 & 0.33 $\pm$ 0.29 \\
 Nonprompt lepton  & 0.44 $\pm$ 0.32 & 2.0 $\pm$ 0.7 & 1.0 $\pm$ 0.6 & $0.03^{+0.14}_{-0.02}$ \\
 Rare  & \NA & $0.14^{+0.39}_{-0.13}$ & \NA & $0.17^{+0.37}_{-0.16}$ \\
 \textbf{Total SM prediction}  & 1.6 $\pm$ 0.7 & 3.5 $\pm$ 0.9 & 2.0 $\pm$ 0.7 & $0.51^{+0.52}_{-0.50}$ \\[\cmsTabSkip]
 Data  & 1 & 0 & 3 & 1 \\  \hline
\\
                        &  \multicolumn{4}{c}{ $\ptmiss>250$\GeV } \\
                        &  $4<\Mll<9$ & $10.5<\Mll<20$ & $20<\Mll<30$  & $30<\Mll<50$                                       \\ \hline
 $\ttbar(2\ell)$  & \NA & 0.19 $\pm$ 0.14 & 0.091 $\pm$ 0.091 & 0.27 $\pm$ 0.14 \\
 DY+jets  & 0.24 $\pm$ 0.19 & 0.24 $\pm$ 0.17 & 0.17 $\pm$ 0.16 & $0.014^{+0.019}_{-0.013}$ \\
 VV  & 0.43 $\pm$ 0.35 & $0.29^{+0.29}_{-0.28}$ & 0.41 $\pm$ 0.29 & 0.66 $\pm$ 0.45 \\
 Nonprompt lepton  & $0.28^{+0.33}_{-0.27}$ & 0.77 $\pm$ 0.44 & 0.38 $\pm$ 0.30 & 0.23 $\pm$ 0.18 \\
 Rare  & $0.45^{+0.57}_{-0.44}$ & \NA & $0.49^{+0.62}_{-0.48}$ & $0.04^{+0.28}_{-0.03}$\\
 \textbf{Total SM prediction}  & 1.4 $\pm$ 0.7 & 1.5 $\pm$ 0.6 & 1.5 $\pm$ 0.8 & 1.2 $\pm$ 0.6 \\[\cmsTabSkip]
 Data  & 2 & 1 & 2 & 0 \\  \hline
\end{tabular}
}
\end{table*}

\begin{table*}[h!t]
\centering
\topcaption{ The number of events observed in
the data and the result of the fit of the backgrounds to the data in the
$\PSQt$ search regions.  The uncertainty indicated is determined from the fit to the 33.2 and 35.9\fbinv integrated
luminosities.  Values for the $\pt(\ell_{1})$ ranges are in \GeV. Rare background event yields are omitted when they do not contribute to the SR bin.\label{tab:PREDh_SR_2}}
\begin{tabular}{l c c c }
                     &  \multicolumn{3}{c}{ $125<\ptmiss<200$\GeV } \\
                     &  $5<\pt (\ell_{1})<12$ & $12<\pt (\ell_{1})<20$ & $20<\pt (\ell_{1})<30$    \\ \hline
$\ttbar(2\ell)$  & 1.9 $\pm$ 0.4 & 11.0 $\pm$ 1.9 & 23.0 $\pm$ 3.5 \\
 DY+jets  & 2.9 $\pm$ 1.4 & 5.6 $\pm$ 1.9 & 4.6 $\pm$ 1.7 \\
  VV  & 0.8 $\pm$ 0.7 & $4.9^{+6.3}_{-4.8}$ & 9.4 $\pm$ 5.4 \\
 Nonprompt lepton  & 8.5 $\pm$ 1.9 & 15.0 $\pm$ 2.6 & 15.0 $\pm$ 2.6 \\
 Rare  & $0.10^{+0.16}_{-0.09}$ & $0.93^{+1.0}_{-0.92}$ & 1.8 $\pm$ 1.7 \\
 \textbf{Total SM prediction}  & 14.0 $\pm$ 2.3 & 37.0 $\pm$ 6.8 & 54.0 $\pm$ 6.5 \\[\cmsTabSkip]
 Data  & 16 & 51 & 67 \\  \hline
\\
                     &  \multicolumn{3}{c}{ $200<\ptmiss<300$\GeV } \\
                     &  $5<\pt (\ell_{1})<12$ & $12<\pt (\ell_{1})<20$ & $20<\pt (\ell_{1})<30$    \\ \hline
 $\ttbar(2\ell)$  & 1.3 $\pm$ 0.35 & 9.9 $\pm$ 1.2 & 15 $\pm$ 2.2 \\
 DY+jets  & 0.92 $\pm$ 0.83 & 2.4 $\pm$ 0.9 & 1.6 $\pm$ 0.6 \\
 VV  & 2.5 $\pm$ 1.4 & 7.1 $\pm$ 4.0 & 12.0 $\pm$ 6.2 \\
 Nonprompt lepton  & 18.0 $\pm$ 3.2 & 20.0 $\pm$ 3.4 & 15.0 $\pm$ 2.7 \\
 Rare  & $0.52^{+0.54}_{-0.51}$ & 1.96 $\pm$ 1.46 & 1.45 $\pm$ 1.13 \\
 \textbf{Total SM prediction}  & 23.0 $\pm$ 3.5 & 41.0 $\pm$ 5.6 & 45.0 $\pm$ 7.0 \\[\cmsTabSkip]
 Data  & 23 & 40 & 44 \\  \hline
\\
                     &  \multicolumn{3}{c}{ $\ptmiss>300$\GeV } \\
                     &  $5<\pt (\ell_{1})<12$ & $12<\pt (\ell_{1})<20$ & $20<\pt (\ell_{1})<30$    \\ \hline
 $\ttbar(2\ell)$  & 0.39 $\pm$ 0.25 & 1.6 $\pm$ 0.5 & 1.6 $\pm$ 0.4 \\
 DY+jets  & 0.33 $\pm$ 0.26 & 0.28 $\pm$ 0.18 & 0.19 $\pm$ 0.07 \\
 VV  & 0.93 $\pm$ 0.53 & 2.5 $\pm$ 1.4 & 4.2 $\pm$ 2.2 \\
 Nonprompt lepton  & 3.1 $\pm$ 1.1 & 5.6 $\pm$ 1.3 & 4.0 $\pm$ 1.3 \\
 Rare  & \NA & $0.15^{+0.18}_{-0.14}$ & $0.45^{+0.50}_{-0.44}$ \\
 \textbf{Total SM prediction}  & 4.7 $\pm$ 1.3 & 10.0 $\pm$ 1.9 & 10.0 $\pm$ 2.5 \\[\cmsTabSkip]
 Data  & 4 & 11 & 9 \\  \hline
\end{tabular}
\end{table*}

\section{Interpretation}

The results are interpreted in terms of the simplified models with compressed
mass spectra for
$\chiz_2\chipm_1\to \cPZ^{*}\PW^{\pm *}\chiz_1\chiz_1$ and for
$\sTop\sTop\to \cPqb\chipm_1 \cPqb\widetilde{\chi}_1^{\mp}$ with the subsequent
decay $\chipm_1 \to \PW^{\pm *} \chiz_1 $ as
discussed in Section~\ref{sec:sample}.
A binned likelihood fit of signal and the background expectations to the data is performed.
This fit takes as input the yields in the SRs (12 for the
electroweakino interpretation and 9 for the top squark interpretation), together
with those in the two CRs ($125<\ptmiss<200\GeV$ and
$\ptmiss>200\GeV$) for the \ttbar and DY+jets estimates, and the three \pt bins
for same-sign leptons for the
$\ptmiss>200 \GeV$ CR. These background-dominated bins also help to constrain
the uncertainties in the background taken from simulation and the one predicted by the ``tight-to-loose''
method.

\begin{figure}[!hbt]
\centering
\includegraphics[width=\cmsFigWidth]{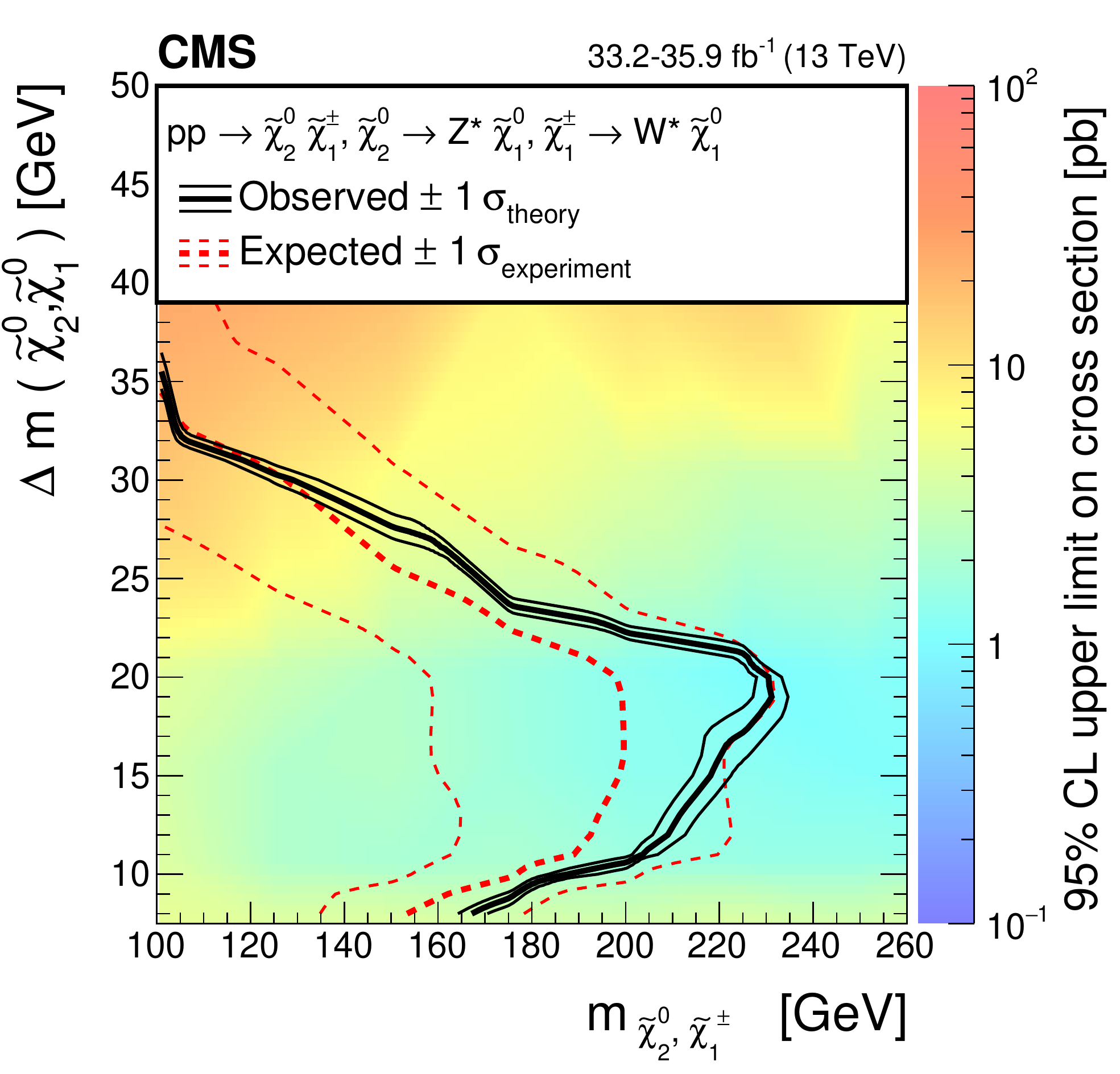}
\caption{The observed 95\% CL exclusion contours (black curves) assuming the
  NLO+NLL cross sections, with the variations corresponding to the uncertainty
  in the cross section for electroweakino. The dashed (red) curves present the
  95\% CL expected limits with the band covering 68\% of the limits in the absence of signal.
  Results are based on a simplified model of
  $\chiz_2 \chipm_1 \to \cPZ^{*}\PW^{*}\chiz_1\chiz_1$ process with a pure
  wino production cross section.}
\label{fig:limits}
\end{figure}

\begin{figure}[!hbt]
\centering
\includegraphics[width=\cmsFigWidth]{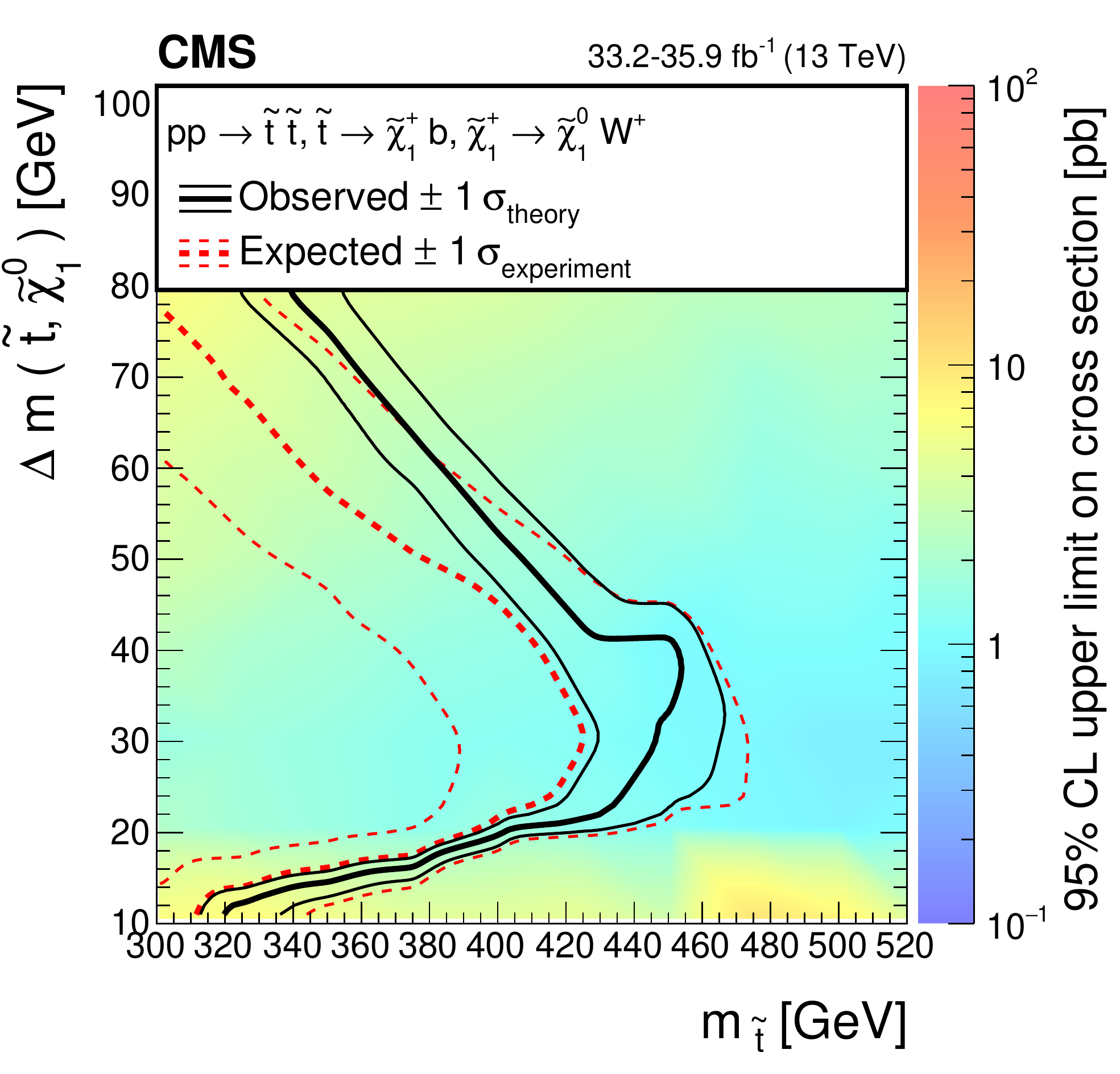}
\caption{The observed 95\% CL exclusion contours (black curves) assuming the
  NLO+NLL cross sections, with the variations corresponding to the uncertainty
  in the cross section for \sTop. The dashed (red) curves present the
  95\% CL expected limits with the band covering 68\% of the limits in the absence of signal.
  A simplified model of the \sTop pair
  production, followed by the $\sTop\to \cPqb\chipm_1$ and the
  subsequent $ \chipm_1 \to  \PW^{*}\chiz_1$ decay is used for the \sTop
  search. In this latter model, the mass of the $ \chipm_1$ is set to be
  $( m_{\sTop} + m_{\chiz_1})/2$.}
\label{fig:limits2}
\end{figure}

Upper limits on the cross sections in the benchmark models at 95\% confidence
level (CL) are extracted. We use asymptotic formulae~\cite{Cowan:2010js} to
derive the results. To set limits, the CL$_\mathrm{s}$ criterion, as described
in~\cite{Junk1999,bib-cls}, is used. Figure~\ref{fig:limits} and~\ref{fig:limits2} show the observed
and expected upper limits on the electroweakino and \sTop pair production cross
sections for the benchmarks considered in this search.

For the electroweakino simplified model, the production cross sections are
computed at NLO+NLL precision in the limit
of a mass degenerate wino $\chiz_2$ and $\chipm_1$, a light bino $\chiz_1$, and assuming
all other SUSY particles
to be heavy and decoupled~\cite{Beenakker:1999xh,Fuks:2012qx,Fuks:2013vua}.
Masses of $\chiz_2$ up to 230\GeV for a $\Delta m(\chiz_2,\chiz_1)$ of 20\GeV are excluded.
The existence of \sTop masses up to 450\GeV
with a $\Delta m (\sTop,\chiz_1)$ of 40\GeV is ruled out for this specific model.

\begin{figure}[!hbt]
\centering
\includegraphics[width=\cmsFigWidth]{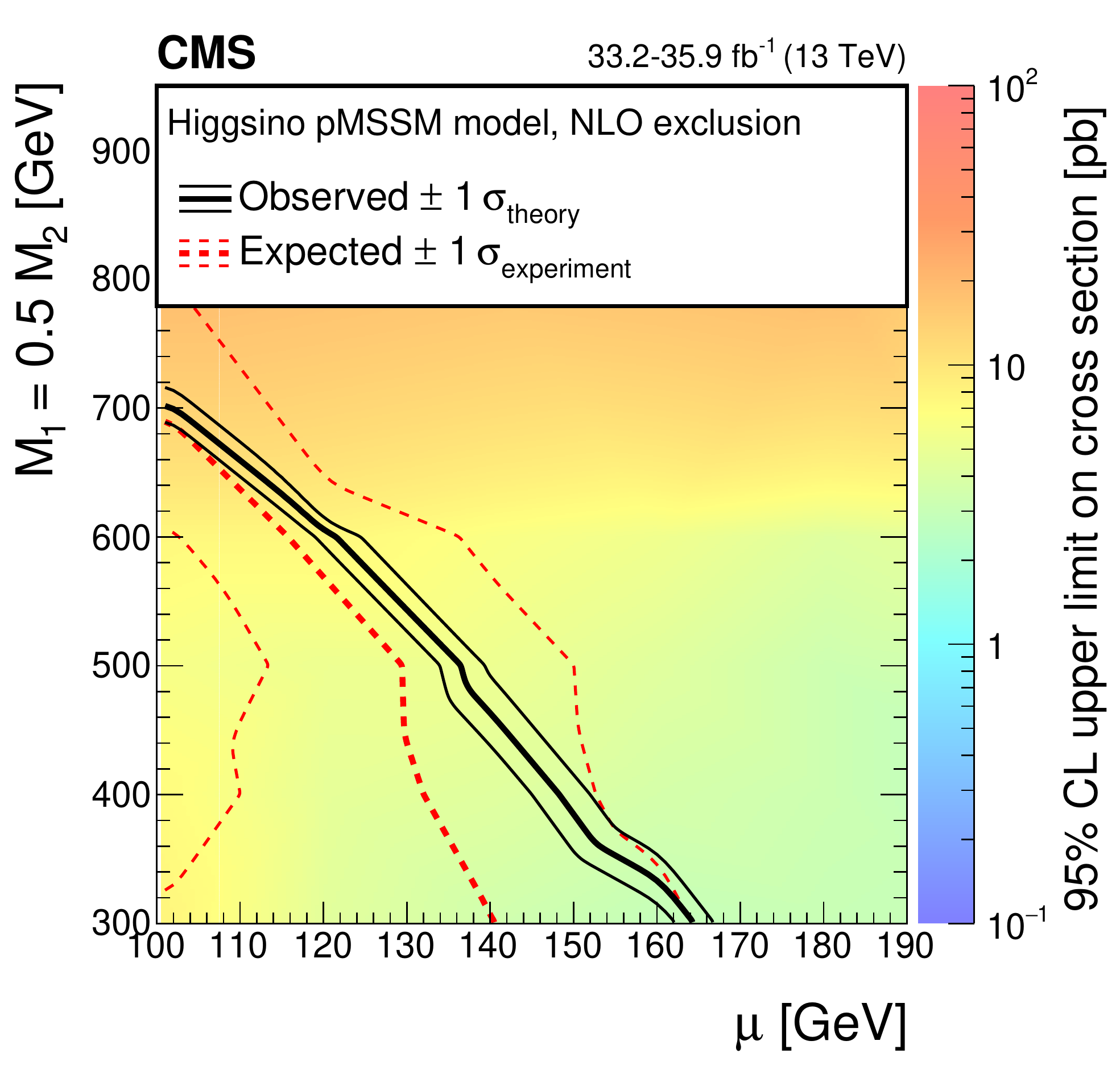}
\caption{The observed 95\% CL exclusion contours (black curve) assuming the NLO
  cross sections, with the variations corresponding to the uncertainty in the cross sections for
  the higgsino pMSSM, which has been introduced in the text. The dashed (red) curves present the band covering 68\% of the limits in the absence of signal.
The model considers all possible production processes.}
\label{fig:limits_higgsinopMSSM}
\end{figure}

The expected and observed exclusion contours for the higgsino pMSSM are shown in
Fig.~\ref{fig:limits_higgsinopMSSM}. The higgsino mass parameter $\mu$ is
excluded up to 160\GeV, when the bino mass parameter $M_1$ is 300\GeV and the
wino mass parameter $M_2$ is 600\GeV. For larger values of $M_1$ and $M_2$, the
mass splitting $\Delta m (\chiz_2, \chiz_1)$ becomes smaller and the sensitivity
is reduced. For $M_1 = 700\GeV$, $\mu$ is excluded up to 100\GeV.

\begin{figure}[!hbtp]
\centering
\includegraphics[width=\cmsFigWidth]{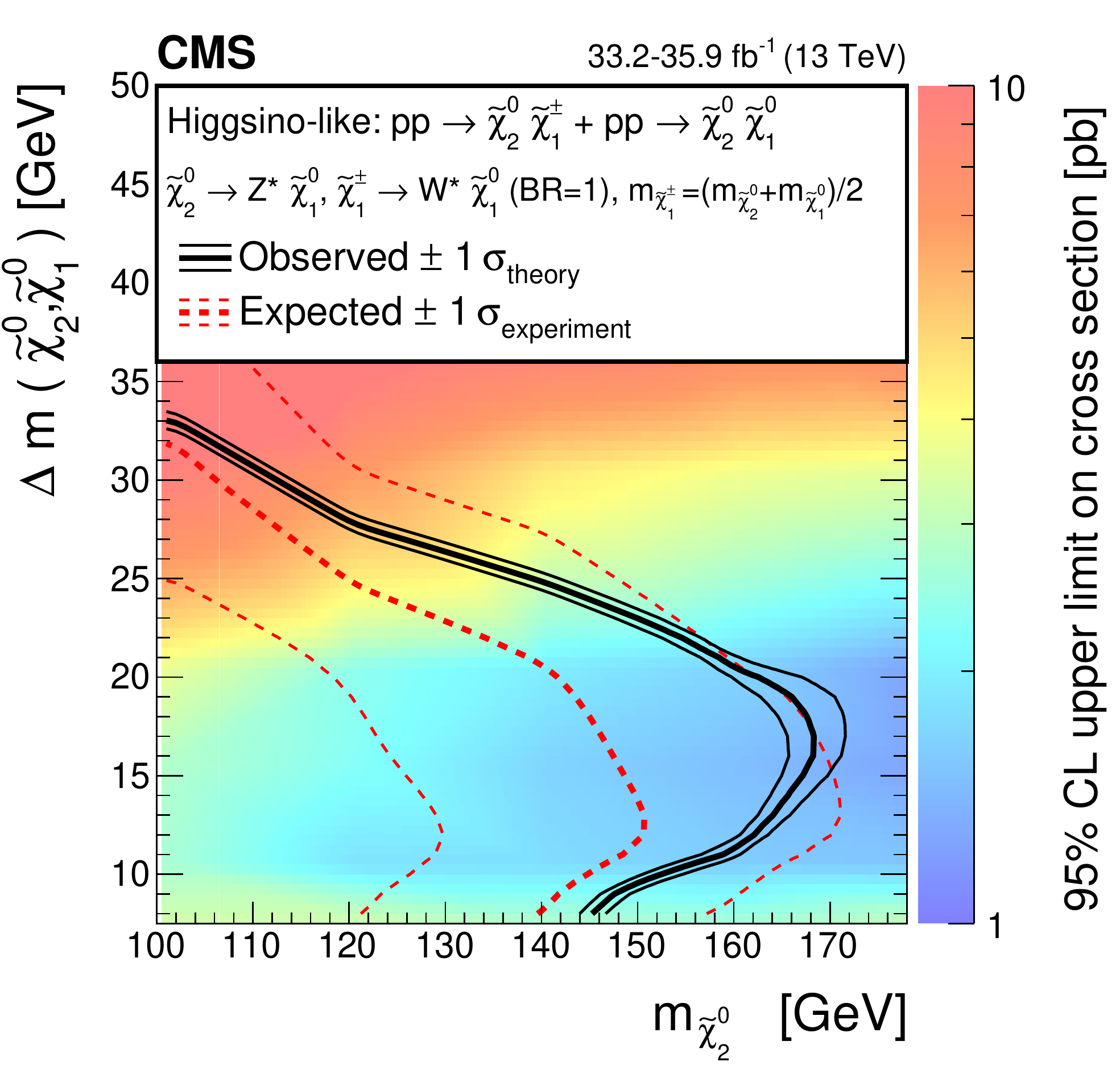}
\caption{The observed 95\% CL exclusion contours (black curves) assuming the NLO+NLL
  cross sections, with the variations corresponding to the uncertainty in the cross sections for
  the higgsino simplified models. The dashed (red) curves present the expected
  limits with the associated band covering 68\% of the limits in the absence of
signal.}
\label{fig:limits_higgsinoSMS_pheno}
\end{figure}

Figure~\ref{fig:limits_higgsinoSMS_pheno} shows the expected and observed exclusion
contours and upper limits on cross sections at 95\% CL in a higgsino simplified model.
To calculate the cross sections in this model, a scan in $\lvert \mu \rvert$, $M_1$, $M_2$ and
$\tan{\beta}$ is carried out. All parameters are required to be real, $M_2$
to be positive and $\tan{\beta} \in [1, 100]$. The remaining SUSY particle masses
are decoupled, and all trilinear couplings are discarded. The parameter space
is then scanned to achieve the maximum higgsino content for $\chiz_2$,
$\chipm_1$, and $\chiz_1$~\cite{Fuks:2017rio}.
For a $\Delta m$ between 15 and 20\GeV, the production model of $\mathrm{pp}\to \chiz_2 \chipm_1$ and $\mathrm{pp}\to\chiz_2 \chiz_1$
is excluded for masses up to $\chiz_2 \sim 167\GeV$.

\ifthenelse{\boolean{cms@external}}{\clearpage}{\clearpage}
\section{Summary}\label{sec:summary}

A search is presented for new physics in events with two low-momentum leptons of
opposite charge and missing transverse momentum in data collected by the CMS
experiment at a centre-of-mass energy of 13\TeV, corresponding to an integrated
luminosity of up to 35.9\fbinv. The data are found to be consistent with
standard model expectations. The results are interpreted in the framework of
supersymmetric simplified models targeting electroweakino mass-degenerate
spectra and \sTop-$\chiz_1$ mass-degenerate benchmark models.  For the \sTop
chargino-mediated decay into $\cPqb\PW^{*}\chiz_1$, top squark masses of up to
450\GeV are excluded in a simplified model for $\Delta m (\sTop,\chiz_1) =
40\GeV$.  The search further probes the $\chiz_2\chipm_1\to
\cPZ^{*}\PW^{*}\chiz_1\chiz_1$ process for mass differences ($\Delta m$) between
$\chiz_2$ and $\chiz_1$ of less than 20\GeV. Assuming wino production cross
sections, $\chiz_2$ masses up to 230\GeV are excluded for $\Delta m$ of 20\GeV.
The search is also sensitive to higgsino production; in a simplified higgsino
model, $\chiz_2$ masses up to 167\GeV are excluded for $\Delta m$ of 15\GeV,
while in a higgsino pMSSM, limits in the higgsino-bino mass parameters
$\mu$-$M_1$ plane are extracted.
\begin{acknowledgments}
We congratulate our colleagues in the CERN accelerator departments for the excellent performance of the LHC and thank the technical and administrative staffs at CERN and at other CMS institutes for their contributions to the success of the CMS effort. In addition, we gratefully acknowledge the computing centres and personnel of the Worldwide LHC Computing Grid for delivering so effectively the computing infrastructure essential to our analyses. Finally, we acknowledge the enduring support for the construction and operation of the LHC and the CMS detector provided by the following funding agencies: BMWFW and FWF (Austria); FNRS and FWO (Belgium); CNPq, CAPES, FAPERJ, and FAPESP (Brazil); MES (Bulgaria); CERN; CAS, MoST, and NSFC (China); COLCIENCIAS (Colombia); MSES and CSF (Croatia); RPF (Cyprus); SENESCYT (Ecuador); MoER, ERC IUT, and ERDF (Estonia); Academy of Finland, MEC, and HIP (Finland); CEA and CNRS/IN2P3 (France); BMBF, DFG, and HGF (Germany); GSRT (Greece); OTKA and NIH (Hungary); DAE and DST (India); IPM (Iran); SFI (Ireland); INFN (Italy); MSIP and NRF (Republic of Korea); LAS (Lithuania); MOE and UM (Malaysia); BUAP, CINVESTAV, CONACYT, LNS, SEP, and UASLP-FAI (Mexico); MBIE (New Zealand); PAEC (Pakistan); MSHE and NSC (Poland); FCT (Portugal); JINR (Dubna); MON, RosAtom, RAS, RFBR and RAEP (Russia); MESTD (Serbia); SEIDI, CPAN, PCTI and FEDER (Spain); Swiss Funding Agencies (Switzerland); MST (Taipei); ThEPCenter, IPST, STAR, and NSTDA (Thailand); TUBITAK and TAEK (Turkey); NASU and SFFR (Ukraine); STFC (United Kingdom); DOE and NSF (USA).

\hyphenation{Rachada-pisek} Individuals have received support from the Marie-Curie programme and the European Research Council and Horizon 2020 Grant, contract No. 675440 (European Union); the Leventis Foundation; the A. P. Sloan Foundation; the Alexander von Humboldt Foundation; the Belgian Federal Science Policy Office; the Fonds pour la Formation \`a la Recherche dans l'Industrie et dans l'Agriculture (FRIA-Belgium); the Agentschap voor Innovatie door Wetenschap en Technologie (IWT-Belgium); the Ministry of Education, Youth and Sports (MEYS) of the Czech Republic; the Council of Science and Industrial Research, India; the HOMING PLUS programme of the Foundation for Polish Science, cofinanced from European Union, Regional Development Fund, the Mobility Plus programme of the Ministry of Science and Higher Education, the National Science Center (Poland), contracts Harmonia 2014/14/M/ST2/00428, Opus 2014/13/B/ST2/02543, 2014/15/B/ST2/03998, and 2015/19/B/ST2/02861, Sonata-bis 2012/07/E/ST2/01406; the National Priorities Research Program by Qatar National Research Fund; the Programa Severo Ochoa del Principado de Asturias; the Thalis and Aristeia programmes cofinanced by EU-ESF and the Greek NSRF; the Rachadapisek Sompot Fund for Postdoctoral Fellowship, Chulalongkorn University and the Chulalongkorn Academic into Its 2nd Century Project Advancement Project (Thailand); the Welch Foundation, contract C-1845; and the Weston Havens Foundation (USA).
\end{acknowledgments}

\bibliography{auto_generated}

\cleardoublepage \appendix\section{The CMS Collaboration \label{app:collab}}\begin{sloppypar}\hyphenpenalty=5000\widowpenalty=500\clubpenalty=5000\vskip\cmsinstskip
\textbf{Yerevan Physics Institute,  Yerevan,  Armenia}\\*[0pt]
A.M.~Sirunyan,  A.~Tumasyan
\vskip\cmsinstskip
\textbf{Institut f\"{u}r Hochenergiephysik,  Wien,  Austria}\\*[0pt]
W.~Adam,  F.~Ambrogi,  E.~Asilar,  T.~Bergauer,  J.~Brandstetter,  E.~Brondolin,  M.~Dragicevic,  J.~Er\"{o},  A.~Escalante Del Valle,  M.~Flechl,  M.~Friedl,  R.~Fr\"{u}hwirth\cmsAuthorMark{1},  V.M.~Ghete,  J.~Grossmann,  J.~Hrubec,  M.~Jeitler\cmsAuthorMark{1},  A.~K\"{o}nig,  N.~Krammer,  I.~Kr\"{a}tschmer,  D.~Liko,  T.~Madlener,  I.~Mikulec,  E.~Pree,  N.~Rad,  H.~Rohringer,  J.~Schieck\cmsAuthorMark{1},  R.~Sch\"{o}fbeck,  M.~Spanring,  D.~Spitzbart,  A.~Taurok,  W.~Waltenberger,  J.~Wittmann,  C.-E.~Wulz\cmsAuthorMark{1},  M.~Zarucki
% ...
\vskip\cmsinstskip
% ...
1:~Also at Vienna University of Technology,  Vienna,  Austria\\
% ...
\end{sloppypar}
\end{document}